\documentclass[10pt,sigconf,letterpaper,nonacm]{acmart}

\usepackage{tikz}
\usepackage{amsmath}
\usepackage{booktabs}
\usepackage{caption}
\usepackage{graphicx}
\usepackage{subfigure} 
\usepackage{makecell}
\usepackage{filecontents}
\usepackage{adjustbox}
\usepackage{multirow}
\usepackage{xspace}
\usepackage{siunitx}
\usepackage{listings}
\usepackage{hyperref}
\usepackage{listings}
\usepackage{xcolor}
\usepackage{courier} 

\definecolor{keywordcolor}{rgb}{0.5,0,0.5}
\definecolor{stringcolor}{rgb}{0.2,0.6,0.2}
\definecolor{commentcolor}{rgb}{0.5,0.5,0.5}
\definecolor{backgroundcolor}{rgb}{0.95,0.95,0.92}

\setcopyright{none}
\makeatletter
\let\@authorsaddresses\@empty
\makeatother
\renewcommand\footnotetextcopyrightpermission[1]{}
\usepackage{color}
\definecolor{lightgray}{rgb}{.9,.9,.9}
\definecolor{darkgray}{rgb}{.4,.4,.4}
\definecolor{purple}{rgb}{0.65, 0.12, 0.82}

\lstdefinelanguage{JavaScript}{
  keywords={typeof, new, true, false, catch, function, return, 
            null, catch, switch, var, if, in, while, 
            do, else, case, break,
            contract, uint256, require, address, external, 
            memory, uint, for, emit, virtual, private, event,
            uint64},
  keywordstyle=\color{keywordcolor}\bfseries,
  ndkeywords={class, export, boolean, throw, implements, import, this},
  ndkeywordstyle=\color{darkgray}\bfseries,
  identifierstyle=\color{black},
  sensitive=false,
  comment=[l]{//},
  morecomment=[s]{/*}{*/},
  commentstyle=\color{commentcolor}\ttfamily,
  stringstyle=\color{stringcolor}\ttfamily,
  morestring=[b]',
  morestring=[b]"
}

\newenvironment{CompactItemize}%
  {\begin{list}{$\triangleright$}%
    {\leftmargin=\parindent \itemsep=2pt \topsep=2pt
     \parsep=0pt \partopsep=0pt}}%
  {\end{list}}

\usepackage{colortbl} 

\newcommand{\offlinetool}{CrossChecked\xspace} 

\newcommand{\hash}[1]{\texttt{\zz#1\zz}}
\def\zz#1{%
 \ifx\zz#1\else
   #1\linebreak[1]\expandafter\zz
 \fi}

\newcommand{\bil}{\textsc{b}\xspace}
\newcommand{\mil}{\textsc{m}\xspace}
\newcommand{\thou}{\textsc{k}\xspace}

\begin{document}
\title{Count of Monte Crypto: Accounting-based Defenses for Cross-Chain Token Bridges} 

\author{
{\rm Enze Liu, Elisa Luo, Jian Chen Yan, Katherine Izhikevich, Stewart Grant,\\ Deian Stefan, Geoffrey M Voelker, Stefan Savage}\\
UC San Diego
}

\iffalse
\newcommand{\alex}[1]{\textcolor{red}{\noindent[AL: #1]}}
\newcommand{\elisa}[1]{\textcolor{blue}{\noindent[EL: #1]}}
\newcommand{\geoff}[1]{\textcolor{teal}{[GV: #1]}}
\newcommand{\deian}[1]{\textcolor{green}{[ds: #1]}}
\else
\newcommand{\alex}[1]{}
\newcommand{\elisa}[1]{}
\newcommand{\geoff}[1]{}
\newcommand{\deian}[1]{}
\fi

\begin{abstract}
Between 2021 and 2023, crypto assets valued at over \$US2.6 billion
were stolen via attacks on ``token bridges'' --- decentralized services
designed to enable the transfer of a token between blockchains.  While the individual
exploits in each attack vary, a single design flaw underlies them all:
the lack of end-to-end value accounting in cross-chain transactions.
In this paper, we empirically analyze over 20 million transactions used
by key token bridges during this period.  We show that a simple invariant
that balances cross-chain inflows and outflows is compatible with
legitimate use, yet precisely identifies every known attack (and
several likely attacks) in this data.  Further, we show that this
approach is not only sufficient for post-hoc audits, but can be
implemented in-line in existing bridge designs to provide generic
protection against a broad array of bridge vulnerabilities. 
\end{abstract}

\maketitle

\section{Introduction}

Careful accounting is key to the correct management of assets in
virtually all financial systems.  Indeed, since Pacioli popularized
double-entry bookkeeping in the 14th century, it has been standard
practice to separately track inflows (credits) and outflows (debits)
to establish a net position --- a balance sheet.

Such accounting is implicit in blockchains as each and every
transaction is recorded, immutable and in possession of implicit
integrity.  Any change in ownership for a given token is explicitly
recorded via some past transaction and thus the net position for
assets in a blockchain is generally consistent and
well-known. However, these integral properties only hold within a
single blockchain.  As soon as one wishes to engage in transactions
\emph{between} chains (e.g., trading Ethereum on Solana), it
requires building a system that steps outside these isolated
environments and implements its own financial calculus between them,
including its own accounting.

Today, one of the principal mechanisms for such transactions is the
cross-chain token bridge --- a service using a pair of ``smart contracts''
(immutable programs stored on a blockchain) to synthesize inter-chain
transactions that are not possible to express natively on a single
chain.\footnote{Some newer blockchain ecosystems, such as Cosmos and
  Avalanche, are themselves composed of multiple blockchains and offer
  some support for bridging across their \emph{own} chains, but not with external chains.}  Such bridges are an extremely popular component of what is
commonly referred to as ``decentralized finance'' (DeFi), and they
routinely manage transaction volumes with value equivalent to \$US8--10
billion per month~\cite{defillama-volume}.

However, token bridges are just code.  They can have bugs in their
implementations, in the other services they depend upon, or in the
mechanisms used to guard the integrity of their cryptographic secrets.
As a result, attackers can, and do, exploit such vulnerabilities to
extract significant value from such systems.  For example, between
2021 and 2023, crypto assets valued at over \$US2.6 billion were
stolen in an array of attacks on token bridges. 

The scale of these hacks has motivated a range of research
focused on automating the detection of such vulnerabilities and
attacks, through the analysis of low-level implementation details
(e.g., using static analysis of smart
contracts~\cite{wang2024xguard,liao2024smartaxe}, predefined anomaly
rules~\cite{zhang2022xscope}, or the features of specific
attacks~\cite{lin2024detecting}).  However, virtually all of these
efforts either require detailed modeling of each contract and bridge infrastructure (e.g.,
the complex protocol bridges run off-chain to relay messages between on-chain contracts),
or are specialized to particular sets of vulnerabilities (e.g., fake deposit events~\cite{lin2024detecting}); they require
continual updating, tuning and re-evaluation as they are tied to
specific implementation artifacts.

In this work, we propose a complementary approach for detecting
attacks on token bridges, by monitoring violations of the \emph{balance invariant}
--- a measure of the difference between value inflows and value
outflows at a token bridge. Unlike prior work, this approach is extremely
simple to reason about because it focuses purely on detecting
potential negative \emph{financial outcomes} instead of focusing on
the precise mechanism of attack that would lead to those outcomes.
Indeed, the power of the approach arises from its technical
simplicity.  Because it abstracts away the complicated implementation
details of bridge contracts (and the off-chain code interfacing with these contracts), the bridge invariant naturally captures 
attacks independent of the details of the vulnerabilities
they target.  We demonstrate this effectiveness by comprehensively
surveying the twelve largest attacks (each \$US1\mil or more) between
2021 and 2023, and showing that while the details of their
vulnerabilities vary considerably, all of them share the property that
transactions are allowed to be ``unbalanced'' (i.e., that outflow can
exceed inflow, less costs).
We then empirically demonstrate, by auditing over 20\mil transactions from 11 different token bridges across 21 blockchains, that auditing transactions for this property is sufficient to retrospectively and automatically identify
each of these past attacks.  

We further show that such audits can be performed \emph{automatically} and in \emph{real-time} --- either to alert bridge operators about potential fraud, or prevent
such unbalanced transactions from completing, thus
preventing any such loss.

We argue that this approach is powerful both due to its simplicity (in a
legitimate financial transaction, the value paid should be equivalent
to the value received, less costs) and its independence from the
vagaries of smart contract details.  Moreover, it represents a
straightforward mechanism that can prevent an entire class of
existing attacks on  bridges.

Our work directly inspired the design and implementation of the Bascule
drawbridge security system, which has been used by Lombard---the largest
Bitcoin
 liquid staking protocol---over the last year to secure the
bridging of
 (over \$1B as of today) BTC deposited on Bitcoin to LBTC on
Ethereum~\cite{bascule}.




\begin{figure*}[h]
\centering
\includegraphics[width=0.8\textwidth]{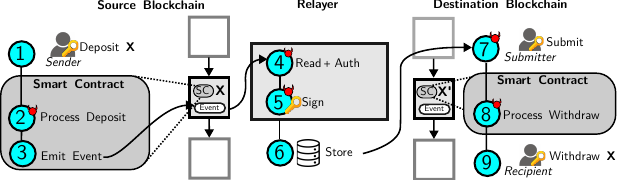}
\caption{Cross-chain token bridging and the steps attackers can exploit to withdraw unbacked deposits.}
\label{fig:cross-chain}
\end{figure*}

\section{Background}
\label{sec:background}
We first give a brief introduction to smart contract blockchains
and cross-chain token bridges. We then describe how token bridges, under attack, collapse in practice.

\subsection{Smart Contract Blockchains} Modern DeFi protocols (e.g.,
decentralized exchanges and lending protocols) are built on top
of ``smart contract'' blockchains, like Ethereum. At their core, these chains, like
the original Bitcoin blockchain, are distributed ledgers that manage accounts
(public keys) and their balances (in native tokens like Ethereum's ETH). Users
interact with these chains by signing and broadcasting (to the distributed nodes
that make up the chain network) transactions that, for example, transfer
funds from their account (using the corresponding account private key) to
another user's account.

Ethereum, and the smart contract chains it inspired since its release in 2015,
differ from Bitcoin by extending the ``simple'' distributed ledger with a smart
contract execution layer. A smart contract on Ethereum is an \emph{internal}
account---and, like a normal, \emph{externally owned} account (EOA), it has a
balance---that has associated \emph{code} (EVM bytecode), which implements the
smart contract's program logic, and \emph{storage}, which persists the program's
state across executions. Users interact with (i.e., execute) smart contracts
much like they do when transferring funds from their account: they sign a
transaction that encodes the smart contract to call (i.e., the contract address),
the particular function to execute, and the arguments to call the function with.
Instead of simply transferring funds from the user's account, then, executing
such a \emph{smart contract call} transaction amounts to executing the smart
contract bytecode---and any smart contracts the contract itself calls.
    





\begin{figure}[t]
    \centering
    \includegraphics[width=\columnwidth]{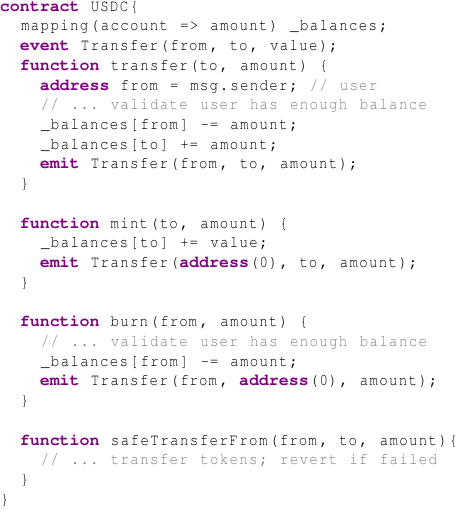}
    \caption{Simplified USDC ERC-20 Token Contract with a Set of Common Functions.}
    \label{fig:erc20}
\end{figure}




\subsection{ERC-20 Tokens}
One of the immediate applications of smart contracts---and to date still one of
most popular---is to create custom tokens (or coins).  To launch a new token, an
organization no longer needs to launch a new blockchain; they can instead deploy
a new contract that implements the ERC-20 token standard interface on a chain
like Ethereum and take advantage of existing on-chain infrastructure like
decentralized exchanges that make it easy to, for example, trade one kind of
token for another---both native tokens (e.g., ETH) and other tokens.\footnote{
    Most EVM chains, i.e., chains that use Ethereum's execution layer,
    follow Ethereum standards like ERC-20. Most non-EVM chains like Solana
    (which have different execution models) have similar standards (e.g., SPL
    tokens in Solana's case).
}

\begin{figure}[t]
\centering
\includegraphics[width=0.5\columnwidth]{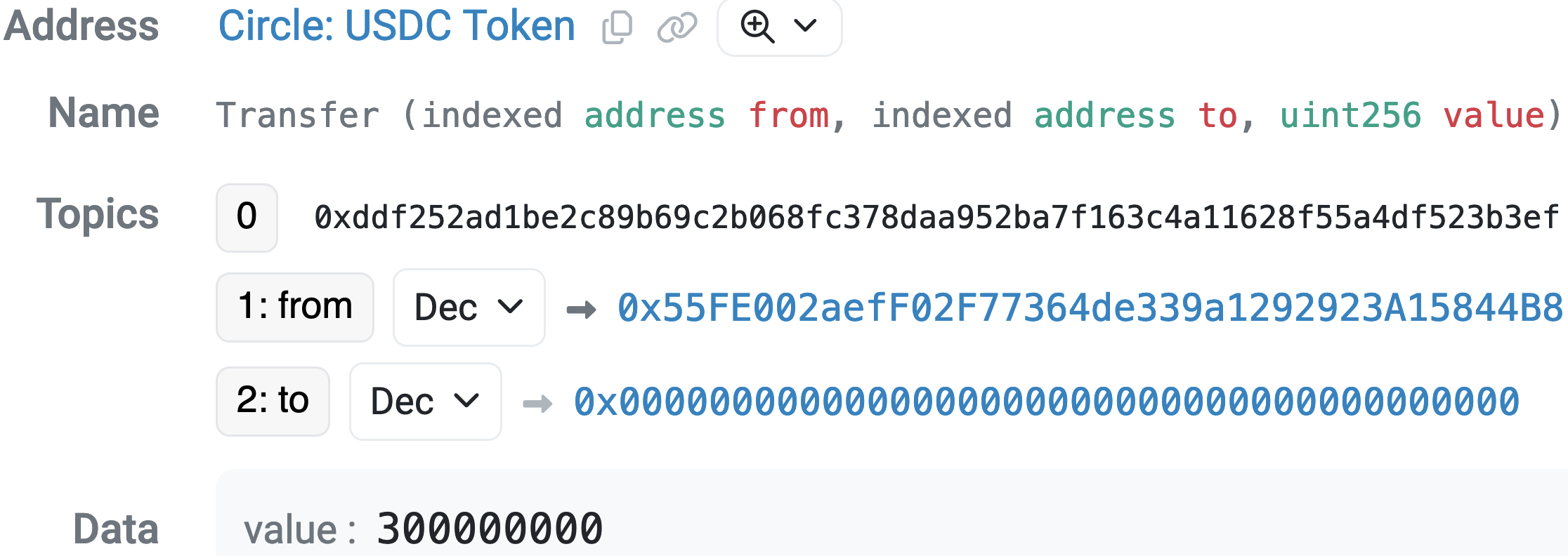}
\caption{Event emitted by an ERC-20 token (USDC).}
\label{fig:erc20-event}
\end{figure}

Figure~\ref{fig:erc20} shows a simplified variant of one such token
contract---the USDC \emph{stablecoin} contract.  This contract tracks
how many USDC tokens an account has and governs the spending of these
tokens (much like a bank governs bank notes).  For example, the
contract's \texttt{mint} function lets Circle (the company that owns
the USDC contract) mint new tokens into a user's account---e.g., after
receiving the corresponding payment from the user off-chain (in US
dollars).  The contract
exposes the ERC-20 interface that lets users (and smart contracts)
transfer tokens from their account by simply calling functions like
\texttt{safeTransferFrom}, and, in turn, use the tokens in any DeFi
protocol (e.g., lending the USDC to different markets, exchanging USDC
for ETH,
etc.). Finally, the contract's
\texttt{burn} function ``burns'' a token out of circulation---and
emits an event (Figure~\ref{fig:erc20-event}) that Circle's off-chain code looks for before allowing
a user to withdraw the corresponding USD fiat off-chain.

\subsection{Cross-Chain Token Bridges}
While smart contracts make it easy to launch new tokens without spinning up new
chains, there is no real shortage of new blockchains being deployed.  Indeed,
the modern blockchain ecosystem is a many-chain ecosystem.  Blockchains like
Avalanche, Base, and Solana have different design points---from cheaper ``gas''
execution costs, to higher throughput, lower latency, and different permission
models---that make them better suited for different classes of application.
This situation has resulted in applications that span many chains and cross-chain
infrastructure that ultimately (try to) allow users to, for example, buy
Dogwifhat NFTs on Solana using USDC tokens on Ethereum.

Core to these applications and infrastructure---and the focus of our work---is
the \emph{cross-chain token bridge}.  At a high level, a cross-chain token
bridge makes it possible for a user to ``transfer'' their tokens from one chain
(e.g., their USDC on Ethereum) to another chain (e.g., Solana) and then use the
transferred tokens on the destination chain (e.g., on a Solana exchange trading
USDC and Dogwifhat).\footnote{While some cross-chain bridges \emph{do} let users
transfer one kind of token (e.g., USDC on Ethereum) for a completely different
kind of token on the destination chain (e.g., Dogwifhat on Solana), these bridges
essentially fuse the cross-chain token bridge with an exchange (or swap). We
focus on token bridges not only because they are fundamental to other kinds of
bridges, but also because they have higher volume, more liquidity, and more
attacks.} Since smart contracts cannot make network requests or otherwise
access state outside their own storage, a cross-chain token bridge consists of
smart contracts on both the source and destination chains, and off-chain
infrastructure that serves to relay the ``transfer'' call across the two
contracts.

As Figure~\ref{fig:cross-chain} shows, a typical cross-chain token transfer,
i.e., a bridge transaction, consists of three phases:

\textbf{1. Deposit (on source chain).}
To initiate a cross-chain token transfer, the user first calls the bridge's
contract on the source chain:
\begin{lstlisting}
deposit(address token, uint256 val, address to) {
  address from = msg.sender;  // user
  address to = address(this); // bridge contract
  // ... validate the transfer request and the ERC-20 token contract
  token.safeTransferFrom(from, to, val);
  emit Deposited(id++, token, from, to, val);
}
\end{lstlisting}
This contract function---a simplified version of 
a bridge deposit
function---processes the user's deposit (step 2 in Figure~\ref{fig:cross-chain}) by validating the transfer request,
e.g., against a list of supported tokens, and then transferring the user's ERC-20 tokens to the bridge contract---recall 
that contracts are accounts with balances.
Then, the function emits an event (step 3 in Figure~\ref{fig:cross-chain}) recording the user's deposit details
(including the deposit ID, the ERC-20 token contract address, the recipient on
the destination chain, and value).

\textbf{2. Off-chain relay.}
The emitted event is observed by an off-chain relayer in (step 4),
which constantly monitors the source blockchain.  The relayer first
verifies the authenticity of the deposit event.  If the event is
authentic, the relayer then produces a signed \emph{receipt}
endorsing the deposit (step 5) and, typically, stores the receipt off-chain (step
6).  Finally, this signed receipt is sent to the bridge's withdrawal contract
on the destination blockchain (step 7). Who submits the receipt varies across
bridges---some bridges submit the receipt on the user's behalf (in these cases, the
signed receipt is simply a signed contract-call transaction), while others give
users (and anyone willing to pay gas) the signed receipt and they, in turn,
submit the receipt to the withdrawal contract to complete the transfer. 

\textbf{3. Withdraw (on destination chain).}
The withdrawal contract on the destination chain first processes the withdraw
request by verifying the receipt then transfers the tokens to the recipient
(step 8). In Chainswap's withdraw case (simplified), for example, users call:
\begin{lstlisting}
withdraw(uint256 id, address token, address to, uint256 val, Signature[] sigs) {
  _chargeFee();
  // verify receipt
  require(received[id][to] == 0, 'withdrawn');
  for(uint i=0; i < sigs.length; i++) {
    verify_receipt(sigs[i], id, token, to, val);
  }
  received[id][to] = val; // mark as withdrawn
  token.safeTransferFrom(address(this), to, val);
  emit Withdraw(id, token, to, val);
}
\end{lstlisting}
This contract function first charges the caller a fee, then verifies the
deposit details against receipt---both that the deposit was not already
withdrawn and that the receipt signatures are valid---and finally transfers the
tokens to the intended recipient.  We consider a cross-chain transaction
complete when the asset is released to the recipient (step~9).

We expect every cross-chain token bridge transaction to uphold the balance invariant:
the value (and kind) of the tokens withdrawn---the outflow---should equal the
value (and kind) of the tokens deposited---the inflow---minus the fees.
In practice, they do not~(Section~\ref{sec:retro-results}).

\section{Threat Model and Attack Vectors}
\label{sec:threat-model}
In this section, we describe our threat model and the attack vectors in scope. 

\textbf{Threat Model}. We consider attacks that seek to steal funds from a 
token bridge. The result of such attacks is that the token bridge is left in an \emph{unbalanced} state, where the outflow of funds (withdrawals) exceeds the inflow of funds (deposits). 
In addition, we assume that the attacker always withdraws funds through the bridge's designated
withdraw function on the destination chain. Otherwise, the attacker can exploit any of the aforementioned components in Figure~\ref{fig:cross-chain} or control the relayer (or submitter) keys. As we show later, our threat model captures the largest attacks on cross-chain bridges that have happened in the past.


\textbf{Attack Vectors}. We consider a broad array of attack vectors: attackers can exploit bugs in all three components---the deposit
contract, the relayer, and the withdraw contract---and use stolen signing keys to steal funds. As we show below, these attack vectors effectively capture attacks that have happened in the past. 
Figure~\ref{fig:cross-chain} highlights the precise steps in a cross-chain transaction that attackers have historically exploited, including:

\begin{CompactItemize}
\item \textbf{Bugs in the deposit contract.} In step 2, the bridge contract
verifies that it has received the correct amount of assets before emitting an
event. Bugs in this verification logic could allow an attacker to deposit a
smaller amount of assets than what is recorded by the bridge in the event (step
3). For example, Qubit's \texttt{deposit} function 
did not properly validate the token address. This bug allowed an attacker to pass \texttt{0} for the token address, so the contract function did not actually transfer any funds from the attacker's account but still emitted a \texttt{Deposit} event which allowed the attacker to withdraw actual tokens on the destination chain~\cite{qubit:rekt}.

\item \textbf{Bugs in the off-chain deposit verification.} In step 5, the
relayer verifies the authenticity of the deposit event emitted by the
bridge contract, including whether the event is emitted by the bridge's
designated contract.  Bridges that do not correctly verify
deposits would allow attackers to withdraw assets that are never deposited.

\item \textbf{Stolen relayer (or submitter) keys.} In step 5, the relayer signs the deposit receipts which are then submitted to the withdrawal
contract as evidence of a valid deposit. If the relayer key is compromised
(e.g., as with the Ronin bridge~\cite{roninattack}) the attacker can forge a
valid receipt and then withdraw assets that were never deposited by calling
\texttt{withdraw} with the forged receipt.

\hspace{0.5cm} The same is true for bridges that submit receipts on behalf of users---and
essentially restrict the \texttt{withdraw} callers to privileged submitter
accounts. The bridge submitter keys (step 8) have similarly been compromised
(e.g., as with AnySwap~\cite{anyswapattack}) and used to withdraw
unbacked deposits.

\item \textbf{Bugs in the withdraw verification.} In step 8, the bridge contract
verifies that the messages are signed by the relayer and have not
been replayed. Bugs in this verification logic (e.g., as we saw with
Wormhole~\cite{wormholeattack}) have allowed attackers to supply ``valid''
payloads that were not signed by the relayer and replay withdrawal
requests with valid deposit receipts that have already been withdrawn.
\end{CompactItemize}

We note that while our threat model captures a wide variety of vulnerabilities, it is not exhaustive. For example, we consider attacks that steal funds from a user (e.g., by tricking them into sending funds to an attacker-controlled address) out of scope. Similarly, we also exclude attacks that do not involve any component of a token bridge or do not have a withdrawal step (e.g., attacks that directly move funds from a token bridge's wallet to an attacker's wallet with a compromised key).
Moreover, we do not consider swap bridges, which introduce additional complexity about the swap logic and the assets being swapped.
Lastly, there can be attacks that withdraw fewer funds than deposited, and we consider these out of scope as bridges 
do not lose funds in such cases and are still in a balanced state.

\section{Approach}

The core hypothesis of our work is that value should be conserved
within cross-chain transactions.  That is, that the value of the asset
inflow in such a transaction (i.e., the deposit) should equal the
value of the asset outflow (i.e., the withdrawal).  In token
bridges, this invariant corresponds to a balancing of the inflow
tokens and the outflow tokens (less any fees or transaction costs
incurred by the bridge itself).  When this balance invariant does not
hold, it allows a range of opportunities for fraud, all of which
involve greater outflows (withdrawals) than inflows
(deposits). 

Testing this invariant on a \emph{per transaction basis} is
straightforward in principle. However, since bridge transaction formats are
not standardized, it requires a range of per-bridge and per-chain
parsing in practice.  Our methodology for normalizing this information
focuses on two key pieces of information: a) identifying each bridge
transaction---a composite of a deposit (inflow) transaction on a
source blockchain and a withdrawal (outflow) transaction on
another---and b) identifying the value transferred in each such bridge
transaction.


\subsection{Identifying and PairingToken Bridge Transactions}
For almost all bridges, identifying their component (per-chain) inflow
and outflow transactions is straightforward --- bridges typically use
explicit events on each chain to signal if a given transaction is a
deposit or withdrawal.\footnote{One key exception to this rule is the
  Wormhole bridge which, until late 2023, did not emit specific
  events when executing a withdrawal transaction.  In this case, we infer that a withdrawal took
  place by looking for transactions that invoke functions designated for performing withdrawal operations (e.g., \textit{completeTransfer}).
  It also appears to be widely understood today that emitting
  explicit events is a best practice.}

Pairing these component transactions (i.e., matching deposits on one
chain to withdrawals on other) can be performed in several different
ways.  The easiest, and most common, is via a unique transaction
identifier. Such IDs are typically generated by the bridge on the
\emph{source chain} during a deposit transaction and then copied into
the withdrawal transaction on the destination chain.\footnote{These
  unique IDs are commonly simple global variables incremented with
  each new transaction.}  In addition, instead of an explicit ID, some
bridges use a hash of the deposit transaction for the same purpose
(e.g., for Anyswap, each withdraw transaction will include the hash of
its corresponding deposit transaction).
Finally, in a handful of cases there are
no ``inband'' identifiers that can be used to associate transactions.
We believe this is a poor design choice that is fundamentally in
conflict with auditability.  However, even in these cases, for the
purpose of our analysis we have been able to pair transactions using
explicit query APIs provided by the affected bridge's
services, which represents the best information available.\footnote{For example, when a bridge transaction on the Poly
  Network bridge includes a withdrawal from Curve (a kind of liquidity
  pool) or when a bridge transaction on the Binance Token Hub includes
  a Binance Smart Chain (BSC) withdrawal, it is not possible to
  identify the partner deposit from blockchain data alone and we must
  make use of ``out of band'' data available through their respective
  bridge query APIs.} 

  At the end of this process, we identify a comprehensive collection
  of ``bridge transactions'': a pair of transactions from two
  different blockchains that were used by the bridge to transfer
  value across them. While the vast majority of bridge transactions
  are pairs of deposit and withdrawal transactions, a handful of
  bridge transactions are either withdrawal-only (e.g., in the case of
  attacks) or deposit-only (e.g., if the user chooses to delay their
  withdrawal).

\subsection{Identifying the Value Transferred}
Prior work~\cite{hu2024piecing} that also studies cross-chain bridges relies on replaying transactions to automatically identify the amount of tokens transferred. However, this approach is not practical for our retrospective analysis, which spans 21 blockchains. For one, some blockchains (e.g., EVMOS) are poorly developed and do not offer easy-to-access replay APIs. Additionally, some transactions cannot be replayed as it is not supported by the blockchain (e.g., Binance Beacon Chain). We show in Section~\ref{sec:live-audit} that 
value extraction can be fully automated when replaying is possible on well-developed blockchains (e.g., Ethereum and BSC).

Instead, for our retrospective analysis, 
we rely on a combination of bridge events and ERC-20 token transfer events, which can be analyzed programmatically. 
In cases where a bridge's transaction event is reliable and explicitly contains the amount and type of tokens transferred, we use that information directly. 
In cases where this is not true, we 
make use of the ERC-20 ``Transfer'' event (see
Figure~\ref{fig:erc20-event} for an example).  This event contains the number of tokens transferred as well
as the sender and recipient addresses, which we use to identify the
number of tokens transferred to or from the bridge. While a Transfer event can also be unreliable in the case of malicious tokens (e.g., tokens that lie in the Transfer event), we have not encountered such cases in our analysis, as bridges typically do not allow arbitrary tokens.  
Because the
Transfer event is adjacent to its associated bridge-generated event,
it is easy to identify and thus establish the number of tokens
transferred.\footnote{As a sanity check, we also require that the
  sender and recipient addresses in the Transfer event are consistent
  with the bridge's defined behavior: for withdrawal transactions, we
  require that the sender is either a mint address (i.e., all-zero
  address) or a bridge-controlled address, while for deposit
  transactions, we require that the recipient is either a burn address
  (i.e., all-zero address) or a bridge-controlled address.} 
  Another special case is caused by so-called ``reflection
tokens'' in which the number of tokens logged in the event (the value
field in Figure~\ref{fig:erc20-event}) is dynamically adjusted based
on a combination of the intended number and the total token supply.
For such cases, we either rely on bridge events which capture the
intended number of tokens transferred or implement token-specific
logic to recompute the value accordingly.  Finally, native tokens have
no Transfer event (since they are not ERC-20 tokens), but thus far we
have either been able to recover the number of tokens transferred from
bridge events or so-called ``internal transactions''.
  
In a few
implementations, multiple related Transfer events can be emitted at
once, and in these cases we have manually inspected their contracts
and constructed implementation-specific logic to account for this
behavior. In total, we had to manually determine the amount of tokens transferred for a few hundred transactions. This manual work was necessary as we could not replay these transactions, and it reflects the limitations in historic data quality, not our methodology. We later show in Section~\ref{sec:live-audit} that such manual work can be avoided entirely when replaying is possible.

To summarize, while it is certainly possible to create a bridge
transaction protocol that records insufficient data to match deposits
and withdrawals, or for which the number of tokens transferred might
be ambiguous, our empirical experience analyzing 11 bridges and 21
blockchains is that such reconstruction has always been possible.

\subsection{Checking the Balance Invariant}
Once we have identified both sides of the bridge transaction and the
number of tokens transferred by the bridge, we can verify if the
balance invariant holds: for each bridge transaction (including those
that are withdrawal-only), does the number of tokens transferred from
the bridge---the withdrawal amount---match the number of tokens
received by the bridge---the deposit amount?

However, a key complication is that bridges can charge fees and, while
these fees can sometimes be paid ``out of band'', it is not uncommon
for them to be subtracted from the tokens received on deposit.  To
account for this behavior (i.e., outflow = inflow - costs), we must be
able to determine such costs on a per-transaction basis.
In most cases, fees can be accounted for in a straightforward manner:
they either are made explicit in bridge-generated events (and can thus
be accounted for directly) or can be calculated based on either
published fee schedules or inferred fee schedules (i.e., since all
transactions are typically subject to the same fixed or percentage
fees).




In some cases, the precise value of fees may be difficult to determine
retrospectively because they depend on some external contemporaneous
value not recorded in the transaction (e.g., fees valued in US dollars
implicitly depend on the exchange rate of a given token at that time).
While such ambiguity could be further minimized with additional data,
in our work we manage this issue by defaulting to a simple rule
that the number of tokens withdrawn should not exceed the number
deposited.


\begin{table*}[t]
  \small
  \centering
  \begin{tabular}{ll}
  \toprule
  \textbf{Blockchain} & \textbf{Bridges that Operate on the Chain} \\
  \midrule
  Arbitrum &      Anyswap, Poly Network, Wormhole \\
  Avalanche &     Anyswap, Poly Network, Meter, Nomad, Wormhole \\
  Binance Beacon & Binance Token Hub \\
  Binance Smart  & Anyswap, Binance Token Hub, Chainswap, Harmony, Meter, Omni, Poly Network, Qubit, Wormhole \\
  Celo &          Anyswap, Poly Network, Wormhole \\
  ETH &           Anyswap, Chainswap, Harmony, HECO, Meter, Nomad, Omni, Poly Network, Qubit, Ronin, Wormhole \\
  EVMOS &         Anyswap, Nomad, Omni \\
  Fantom &        Anyswap, Poly Network, Wormhole\\
  Gnosis &        Omni, Poly Network       \\
  Harmony &       Anyswap, Harmony Bridge, Poly Network \\
  HECO &          Anyswap, Chainswap, HECO, Poly Network \\
  Meter &         Meter\\
  Metis &         Anyswap, Poly Network\\
  Milkomeda &     Nomad \\
  Moonbeam &      Anyswap, Meter, Nomad, Wormhole\\
  Moonriver &     Anyswap, Meter \\
  OKT Chain &     Anyswap, Chainswap, Poly Network \\
  Optimism &      Anyswap, Poly Network, Wormhole\\
  Polygon &       Anyswap, Chainswap, Poly Network, Wormhole \\
  Ronin &         Ronin\\
  Solana &        Wormhole\\
  \bottomrule
\end{tabular}
  \caption{The blockchains \offlinetool supports and the cross-chain
    bridges that operate on them.  While a particular attack on a
    bridge involves two chains, we collect deposit and withdrawal
    transactions for a bridge on all chains that the bridge supports. Anyswap is also known as Multichain.}
\label{table:chain_bridge_full}
\vspace{-0.05cm}
\end{table*}

\begin{table*}[t]
\centering
\begin{tabular}{p{2.8cm}rS[table-format=4.1]rrcrrr}
  \toprule
  \textbf{Name} & \multicolumn{1}{c}{\textbf{Date}} & {\textbf{Loss (USD)}} & \textbf{Analyzed} & \textbf{Reported} & \textbf{New} & \textbf{Test} & \textbf{Error} & \textbf{Suspicious} \\
  \midrule
    Ronin & Mar 2022 & 624.0\mil & 3.0\mil & 2                 & - &  -  & - & - \\ 
    PolyNetwork/2021 & Aug 2021 & 611.0\mil & 292\thou & 18    & - &  -  & 1 & - \\ 
    BSC Token Hub & Oct 2022 & 587.0\mil & 2.0\mil & 2         & - &  -  & - & - \\ 
    Wormhole & Jan 2022 & 360.0\mil & 642\thou & 1             & - &  -  & 2 & - \\ 
    Nomad & Aug 2022 & 152.0\mil & 37\thou & $^\dagger$962      & - &  -  & - & - \\ 
    Harmony & Jun 2022 & 100.0\mil & 336\thou & 15             & - &  -  & - & 43 \\ 
    HECO & Nov 2023 & 86.0\mil & 23\thou & 8                   & - &  -  & - & 73 \\ 
    Qubit & Jan 2022 & 80.0\mil & 260 & 16                     & - & 114 & - & - \\ 
    Anyswap & Jul 2021 & 7.9\mil & 3.4\mil & 4                & 24 &  -  & 240 & 29 \\ 
    PolyNetwork/2023 & Jun 2023 & 4.4\mil & 290\thou & 136     & - &  -  & 1 & 27 \\ 
    Chainswap & Jul 2021 & 4.4\mil & 53\thou & $^*$1136        & - &  15 & 4 & - \\ 
    Meter & Feb 2022 & 4.3\mil & 14\thou & $^\dagger$5          & - &  -  & - & - \\ \midrule 
    Total  &          & 2.6\bil & 10.1\mil & 2,305             & 24 &  129 & 248 & 167 \\
    \bottomrule
\end{tabular}
\caption{List of top attacks on cross-chain bridges in the retrospective
  analysis, ordered by amount stolen.  \offlinetool analyzed over
  10\mil bridge transactions (20\mil component deposit and withdrawal transactions)
  and identified all bridge transactions previously reported as having
  been associated with the attacks (Reported).  It also identified
  bridge transactions that violated the invariant that were a
  previously unidentified attack (New), test transactions (Test),
  transactions reflecting bugs in implementations or use (Error), and
  suspicious transactions that employ manual signing (Suspicious).
  $^\dagger$\offlinetool identified slightly more transactions than
  were reported by the Nomad~\cite{nomad-groundtruth-github:online}
  and Meter~\cite{meter-groundtruth-tencent:online} bridges for their
  attacks (+2 for Nomad, +1 for Meter).
   $^*$The Chainswap attack only has reports of the malicious deposit
  address~\cite{chainswap-groundtruth:online}, which matches the one
  identified by \offlinetool.
  %
}
\label{table:bridges-and-txns-flagged}
\end{table*}

\section{Retrospective Analysis}
\label{sec:retro-results}

The basic reasoning motivating the balance invariant is simple: that
legitimate bridge transactions should conserve value.  However, this
assumes a particular model for how bridges are used and operated which
may or may not hold in practice.  To validate our hypothesis, we applied
the balance invariant analysis retrospectively across a large set of
past transactions which, while primarily benign, contain the largest
known bridge attacks during our period of study.  Our goal is both to
show that all real attacks are identified (detection), but also that
the invariant does not alert on large numbers of benign transactions
(bridge compatibility).  Later, in Sections~\ref{sec:live-audit}
and~\ref{sec:active-protect}, we will describe systems for live bridge
monitoring as a third party and a new implementation for preventing
unbalanced transactions from being committed.



\subsection{Data Set}
\label{sec:retro-data}


To perform a retrospective analysis we need data for which (at
least some) of the ground truth is known.   In this section we
describe how we chose the attacks we study, the blockchains and
smart contracts involved, and how we collected the historical
transaction data.

\textbf{Attack Selection.}
%
We compiled a comprehensive list of attacks on cross-chain token
bridges that occurred between January 2021 and December 2023.  We used both academic papers surveying cross-chain
bridge attacks~\cite{lee2023sok, zhang2023sok, zhao2023comprehensive} as well
as industry blog posts that collect and characterize attacks over time~\cite{GithubBridgeBugTracker, SlowMistHackedBridges:online,
  REKTDB:online, Web3Great:online, GithubBridgeHacks2:online}. 
As described in our threat model (Section~\ref{sec:threat-model}), we filter out attacks
involving blockchains lacking smart contracts such
as Bitcoin (e.g., the pNetwork attack in 2021~\cite{pNetworkhack:online}), attacks on swap bridges (e.g., the Thor bridge attacks~\cite{Thorhack1:online,Thorhack2:online}), attacks that do not involve individual bridge transactions (e.g., the evoDefi attack~\cite{evoDefihack:online}),
and attacks that withdraw funds without a bridge component (e.g., direct transfer from a vault as in the Multichain 2023 attack~\cite{Multichainhack:online}).  Moreover, given the large number of attacks during this period, we focus on
major attacks with claimed losses greater than \$1 million USD and
exclude the remaining (the sum of losses from these excluded attacks are a small fraction of those in our scope).
%
%
This process yielded 12 attacks on 11 distinct bridges.
We note that this list
includes the top five attacks on cross-chain bridges in history, which
collectively resulted in more than \$2.6 billion USD in claimed losses.

\textbf{Blockchain Selection.}
%
Validating bridge transactions requires access to
transaction data from both the source and destination chains.  We
support every blockchain involved in the bridge transactions
associated with the 12 attacks, either as the source (or claimed
source) or destination chain.  As a result, we support a total of 21
blockchains that together cover a
broad range of designs, including many of the most popular such as 
Ethereum, Binance Smart Chain, and Solana.
%
Table~\ref{table:chain_bridge_full} lists the blockchains and the
bridges in our retrospective study that operate on them.





\textbf{Smart Contract Selection.}
For each bridge, we comprehensively collected the results of all
versions of its bridging smart contracts on every blockchain we
considered.  In particular, we collected deposit and withdrawal
transactions created by every contract the bridge deployed on
every blockchain we support (typically many more blockchains than the
ones involved in a bridge attack) as well as all versions of the smart
contract implementation for the bridge (bridges evolve their
implementations over time, such as in response to an attack).
Since the transactions we collected from this set of smart contracts
are much broader than those just involved in the top 12 attacks we
consider, including them further reinforces our findings that the
alerting workload is very small (Section~\ref{sec:retro-analysis}).




\begin{figure}[t]
  \centering
  \includegraphics[width=\columnwidth]{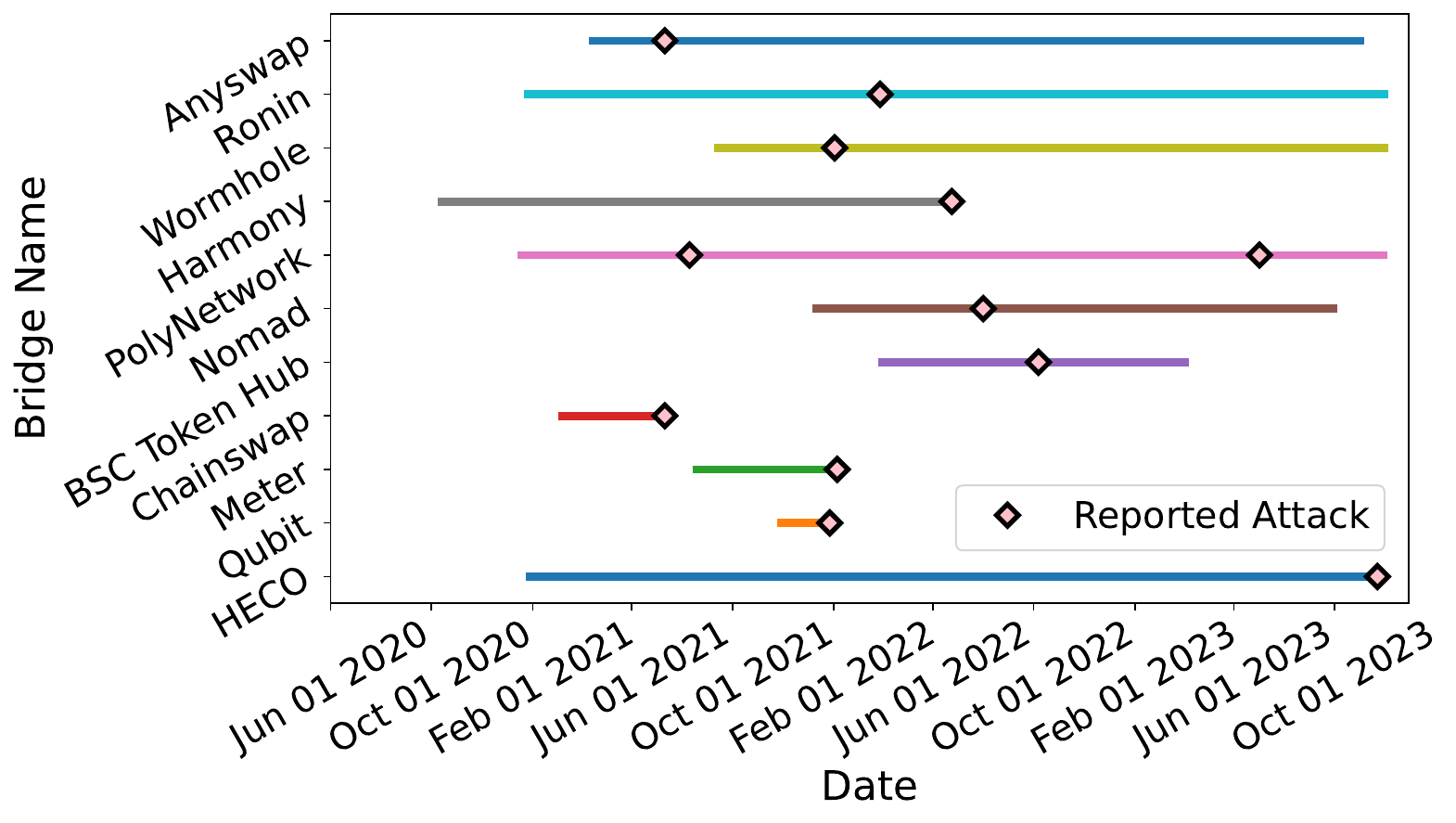}
  \caption{The lifetime of the bridges in our retrospective study.
    Lines start with the bridge's first valid transaction and end with the
    last valid transaction in our data, corresponding to the bridge's
    closure or November 2023 if the bridge was still operating at the
    end of our data set.  Diamonds indicate the dates of attack.}
  \label{fig:bridge-timeline}
\end{figure}

\textbf{Data Collection.}
Collecting smart contract transaction data---the verbose record of
what each contract did---across many chains constitutes the most
time-intensive aspect of performing a retrospective analysis.

For each bridge, we collected deposit and withdrawal transactions for
each of the blockchains it operates on (for the 21 chains we
support) using commercial RPC services which charge for queries.  We
primarily used five commercial services (GetBlock, QuickNode, ChainStack, GoldRush, and Ankr.),\footnote{
GetBlock, QuickNode, ChainStack, GoldRush, and Ankr.} 
as no single service supports
all of the blockchains and functionality we need.
For each bridge and blockchain combination, by default we
collected all of the historical transactions generated by its smart
contracts from the genesis of its deployment to the end of November
2023 or the end of its life, whichever came first.

There are two exceptions to this rule. For the Binance bridge, we
limited data collection to a year (six months prior to its attack and
six months after) because of the sheer volume of transactions over its
lifetime (30 million transactions) and the limited time available before the sunset of the Binance Beacon Chain.
For the Harmony bridge, whose
contract was reused by another bridge (LayerZero) after its attack
(since LayerZero was not attacked itself, we only consider transactions that were part of Harmony).

%
%

Figure~\ref{fig:bridge-timeline} illustrates these bridge lifetimes
with each line starting at a given bridge's first transaction and
ending with its last in our data set.  Diamonds indicate the dates of
individual attacks, emphasizing that many of the bridges closed
shortly after their attacks.

\subsection{Analysis}
\label{sec:retro-analysis}


We developed a tool called \offlinetool that pairs deposit and
withdrawal transactions from blockchains into bridge transactions, and
applies the balance invariant and consistency checks on them.\footnote{As consistent with prior work~\cite{hu2024piecing}, we ignore transactions for which we cannot pair because of incomplete information (e.g., deposits to a valid blockchain that we do not support).} Using
the transactions we collected for the 12 attacks across 11 bridges and
21 chains, \offlinetool analyzed over 10\mil bridge transactions (20\mil individual deposit and withdrawal transactions),
identifying more than 2.3\thou bridge transactions associated with the
attacks and 568 more that violated the balance invariant.


Table~\ref{table:bridges-and-txns-flagged} summarizes our results.
For each attack, it shows the bridge involved, the date of the attack,
the claimed loss in USD, the number of transactions \offlinetool
analyzed, and the number and kinds of transactions that violated the
balance invariant.
Below we discuss these two categories of transactions in more detail.
For further context on each of the attacks,
Appendix~\ref{sec:attack-summaries} provides a descriptive account
summarizing how the attacks transpired.

Looking forward to Sections~\ref{sec:live-audit}
and~\ref{sec:active-protect}, altogether \offlinetool identified 2,873
bridge transactions (0.03\%) that violated the balance invariant out
of more than 10\mil bridge transactions analyzed.  If the invariant is used by a
third-party auditing or protection system, we note that such an alert
workload has a negligible overhead for manual inspection, typically
raising no more than one alert (or one batch of alerts) every few
weeks.

\subsubsection{Known}

For each of the 12 significant cross-chain attacks,
Section~\ref{sec:retro-data} describes how we gathered the historical
transactions that correspond to the attacks using external sources.
We use these identified transactions as ground truth for evaluating
the ability of \offlinetool to identify attack transactions.  As shown
in Table~\ref{table:bridges-and-txns-flagged}, when processing the
transaction histories of the chains involved, \offlinetool
successfully identified all 2,305 bridge transactions on the source
and destination chains associated with the attacks (including a few ones that are missed by all public reports).

Since we designed \offlinetool to specifically identify such attacks,
these results may not be surprising.  However, they are useful for
confirming that the approach of checking a simple, well-defined
invariant works well.  Moreover, the approach works well across a
variety of models, including bridges that specify fees in a fiat
currency, tokens that use a reflection mechanism, etc.

\subsubsection{Other Violating Transactions}


Equally compelling
is the extent to which other, non-attack
transactions violate the balance invariants.  If the
attack transactions are dominated by many false positives, then the
approach becomes less effective.
As shown in Table~\ref{table:bridges-and-txns-flagged}, \offlinetool
finds significantly fewer (568 compared to 2,305) bridge transactions that were
not previously identified as attacks.  Given the nature of these
additional transactions, though, they do not undermine the
effectiveness of the invariant approach.  By violating the invariant
something highly unusual is taking place.  As a result, we argue that
such transactions should be flagged for further scrutiny and perhaps
even blocked from executing (particularly transactions in the New category and the large transactions in
the Suspicious category below).

To ensure that we have not missed benign explanations for a violation,
we manually inspected violating bridge transactions in at least one
of the following ways: (1) using blockchain explorers to verify that
the funds have been created by the withdrawal (and if so, whether they
have been moved); (2) searching online for any additional information
about the transaction and addresses involved; (3) checking that if a 
claimed deposit exists; and (4) examining unredeemed
deposits on the source chain that potentially could have been used to
back the withdrawal (e.g., because the smart contract implementation
changed between the deposit and withdrawal).  If we can manually
reconcile a bridge transaction, we consider it benign
and do not consider it further.

We group the remaining bridge transactions that violate the invariant
into four categories, which we describe below.  For reference, we also
list some of these bridge transactions in
Table~\ref{tab:xaction-hashes} in the appendix to provide more detail.


\newcommand{\pgraph}[1]{\vspace*{0.1in}\noindent\textbf{#1}}

\pgraph{New.}  We believe that we have identified
previously unreported bridge transactions involved in two new,
unreported attacks on Anyswap.
The first group of three transactions executed once Anyswap reopened
after the attack but before Anyswap patched its smart contract
(Anyswap transactions \#1--3 in Table~\ref{tab:xaction-hashes} in the
appendix).  These transactions were three days later and involved
different deposit and withdrawal addresses than the July 10, 2021
attack (yet utilizing the same compromised key).  The second group of 21 transactions on November 18, 2021,
were all withdrawing on Avalanche.  These
transactions either referenced deposits that had already been redeemed previously (Anyswap \#6 in Table~\ref{tab:xaction-hashes}) or referenced non-existing deposits (Anyswap \#7 in Table~\ref{tab:xaction-hashes}).  Manually
inspecting the smart contract used, the attackers were exploiting a
bug in Anyswap's implementation (missing access control code) and one of the addresses was labeled as ``KyberSwap Exploiter''. These results also match those reported by Hu et al.~\cite{hu2024piecing}.

\pgraph{Tests.}  Two bridges had test transactions that violated the
invariant.
Chainswap had 15 transactions that withdrew tokens labeled as test
tokens (e.g., tokens labelled as ``testtoken'' or ``startertoken''). Similarly, Qubit had 114 transactions that minted test tokens (e.g., ``xTST'') without backing deposits.
While all these transactions did not have corresponding deposits, we surmise that they were likely benign given the tokens involved (tokens that have no real value).

\pgraph{Error.} Four bridges had transactions that suggest bugs in
either their implementation or their invocation.

PolyNetwork/2021 had one bridge transaction where the withdrawal
amount matched the deposit amount, but the withdrawal moved the funds
to the wrong destination chain (\offlinetool flagged the mismatch in
destination chain specified in the deposit and withdrawal
transactions).
Anyswap had 238 withdrawal transactions that pointed to just a few deposits. Manually sampling a few, the deposit transaction hash
was not set correctly in the withdrawal transactions (we found their matching deposits), suggesting a program error. In addition, Anyswap had two withdrawals referencing deposits that did not specify a recipient, yet the deposit
amounts covered the withdrawals.  While technically not balance
invariant violations, \offlinetool flagged them because of their
unusual circumstances.

%
Wormhole had two bridge transactions that violated the invariant: 0.5
wrapped Solana on Polygon $\rightarrow$ 650 USDC on Solana, and
4.3\thou MATIC on Polygon $\rightarrow$ 2.3\thou on Avalanche.  The
small amounts and close proximity of the dates of the transactions
suggest they are also likely errors.

Finally, three bridges had transactions that had no apparent effect,
suggesting invocation errors or undocumented behaviors.
PolyNetwork/2023 had one bridge transaction, and Chainswap had four,
where the deposits were non-zero but the withdrawal amount was zero. 



\pgraph{Suspicious.}
%
The final category of bridge transactions suggests the manual,
intentional use of private keys in signing transactions that
effectively bypass verification --- precisely the kind of transactions
that warrant auditing.


Many of these cases involved highly suspicious unbacked bridge
transactions involving very large withdrawals without corresponding
deposits.  For example, Anyswap had 29 transactions totaling more than
\$1.5\mil pointing to non-existent or already-redeemed deposits (e.g., Anyswap \#4 and \#5 in
Table~\ref{tab:xaction-hashes} in the appendix).
PolyNetwork/2023 had 27 withdrawals referencing a non-existent
deposit address totaling more than \$20\mil (PolyNetwork/2023 \#1 in Table~\ref{tab:xaction-hashes}).

The HECO bridge had 73 unusual transactions.  One was a very
suspicious unbacked bridge transaction minting \$5\mil of USDT on the
HECO chain (HECO \#1 in Table~\ref{tab:xaction-hashes}).  The
remaining 72, totaling over \$36\mil, involved withdrawals to an
address labeled ``HECO recovery''.  The label suggests benign intent
such as rescuing funds trapped in the bridge, but the activity is also
consistent with a rug pull.


Other cases suggest seemingly careless operational practices.  In particular,
Harmony had 43 bridge transactions that violated the invariant in different ways.
Eight double-spending bridge transactions (e.g., Harmony \#1 in Table~\ref{tab:xaction-hashes}) used the same deposit to
release tokens twice (though only resulting in a profit of a few
hundred USD).  
Thirty-two
had indecipherable data:
it was not possible to decode the deposit function name, function
input, and some events (thus preventing verification of the deposit).
One bridge transaction minted tokens on Harmony
chain referencing a non-existent deposit.  And two
had withdrawal amounts that were smaller than the deposits, perhaps
caused by undocumented fees or errors.
Considering the specific mechanism used by the Harmony bridge, where a privileged submitter key
submitted transactions that it should not have, 
it suggests that Harmony had issues operating securely and
correctly.

\section{Live Auditing}
\label{sec:live-audit}
Section~\ref{sec:retro-results} shows that the balance invariant can accurately identify past bridge attacks with a small amount of manual effort. In this section, we show that this invariant can be used to monitor ongoing transactions in a production bridge in a fully automated fashion.
Since most of the bridges in our retrospective analysis are shut down at
the time of writing (Figure~\ref{fig:bridge-timeline}), we focus on the
Wormhole bridge. 
Wormhole is still operational, supports a wide
variety of blockchains, and is among the most popular bridges in terms
of transaction activity~\cite{zhang2023sok}.  This live system
monitors around 1\thou withdrawal transactions per day across 10
blockchains (Wormhole operates on 10 of the 21 blockchains we support), which roughly account for 40\% of all withdrawal
transactions.  Extending the system to support other
bridges is straightforward.



\subsection{System Overview}

Our system has two main components. 
The
\textit{Blockchain Monitor} tracks the blockchain network and
retrieves the latest deposit and withdrawal transactions. The
\textit{Auditor} checks extracted transactions for invariant
violations. 



\textbf{Blockchain Monitor.}
The Blockchain Monitor obtains deposit and withdrawal transactions by
periodically retrieving the latest finalized blocks.  The Monitor
uses the same commercial RPC services as we used for the retrospective
analysis (Section~\ref{sec:retro-data}).

The Monitor only retrieves finalized blocks since unfinalized blocks
may be reverted or reordered.  Since different blockchains have
different finality times, we use the timestamp of the latest finalized
block of Ethereum as the synchronization time since Ethereum has the
slowest finality time (around 12 minutes)~\cite{Finality:online}.
After retrieving the blocks, the Monitor extracts deposit and
withdrawal transactions for Wormhole and saves them to a local
database. The Monitor polls every minute (configurable) for new finalized blocks.
\textbf{Auditor.}
For every new withdrawal transaction extracted by the Blockchain
Monitor, the Auditor attempts to find the corresponding deposit
transaction in its local database.  In most cases it will find the
deposit, but if it cannot find it (e.g., because the
withdrawal references a non-existent transaction), the Auditor sends
an email alert inviting manual inspection. If it finds a deposit, it
examines the deposit and withdrawal transactions. To compute the amount of tokens transferred, the Auditor replays the transactions, which is completely automated and involves no manual effort.
The auditor sends an
alert if either the deposit has already been withdrawn (the new
withdrawal is double-spending) or the bridge transaction violates
the balance invariant.

\subsection{Deployment}
We have deployed our live auditing system for the Wormhole bridge over 10 months (October 2024 to June 2025).
The system audits the deposit and withdrawal transactions between 10
blockchains, which we estimate account for around 40\% of all
withdrawal transactions on Wormhole. Our system audited over 202\thou transactions (more than 900 transactions per day) and sends email
alerts if it observes a transaction violating the balance invariant or
another auditing property (e.g., the destination chain in the deposit
does not match the chain in the withdrawal).

In the time that our system has been operational, it has alerted on 103 transactions. Upon manual inspection, we confirmed that all alerts were caused by bugs in our code or the RPC services we used (e.g., failing to index a transaction, which typically resolves after retrying after a few minutes). Consistent with our retrospective analysis, the alert rate is very low at 0.05\% and raises less than one alert per day. 

In addition, while we have not observed any attacks during our
monitoring period, we simulated three attack scenarios to confirm that
the system alerts when expected.  These three transactions represent
three types of attacks: (1) a double-spend attack, (2) an
unbacked withdrawal attack, and (3) an attack where the deposit and
withdrawal amounts do not match. The system successfully alerted on
all three. 


\section{Protecting Bridges}
\label{sec:active-protect}

\begin{figure}[t]
\centering
\includegraphics[width=0.9\columnwidth, trim=0 1.2 0 0, clip]{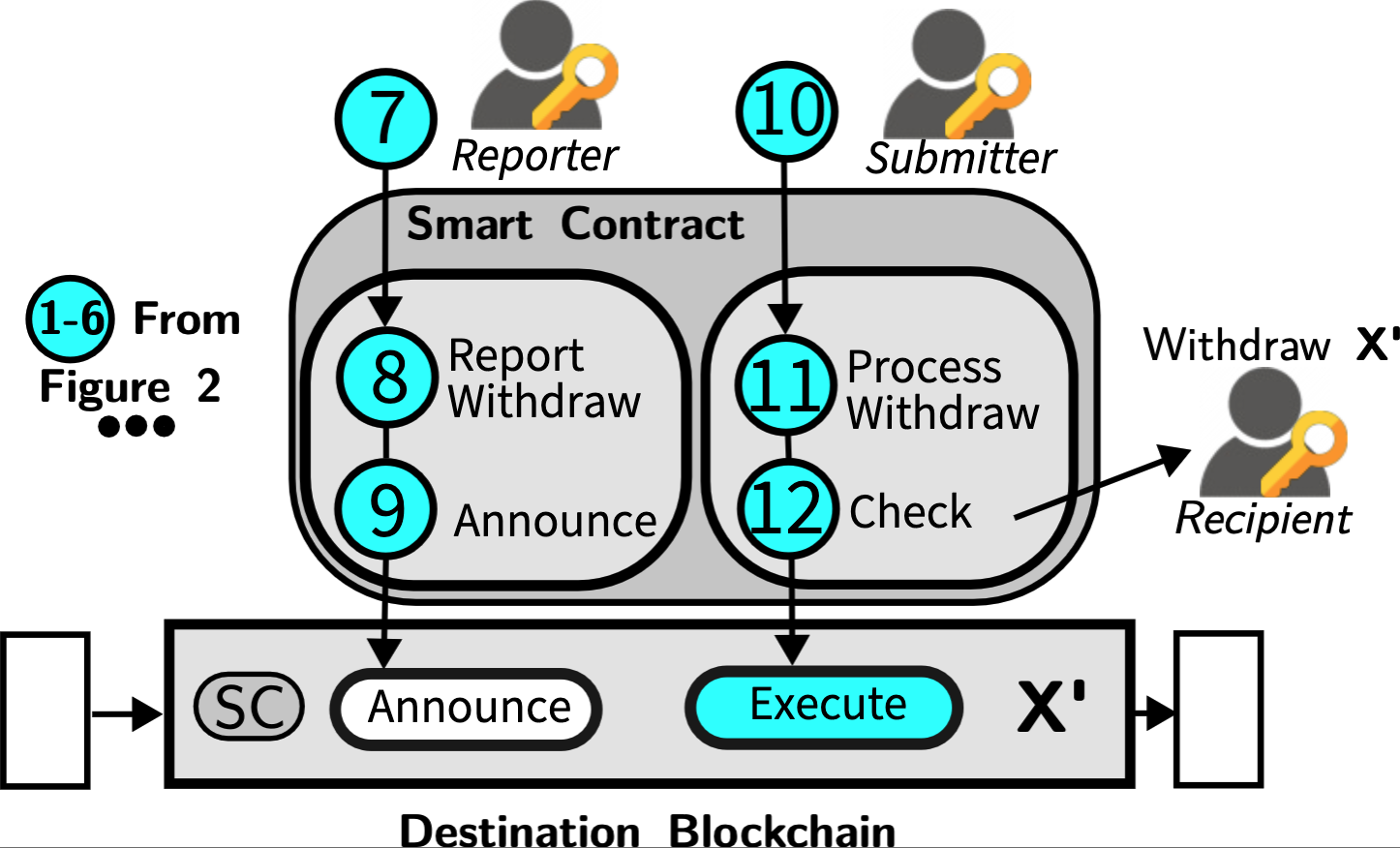}
\caption{Report-then-execute model for bridges. The reporter 
asynchronously reports valid transactions and announces them through events.}
\label{fig:improved-arch}
\vspace{-0.3in}
\end{figure}

Sections~\ref{sec:retro-results} and~\ref{sec:live-audit} showed how
applying the balance invariant could identify attacks retrospectively and
monitor ongoing transactions as an external party without any
changes to existing bridge infrastructure.  However, the live
monitoring system only alerts on violations of the invariant and does
not prevent violating bridge transactions from being processed.

In this section, we describe an approach for extending an existing
bridge implementation to prevent malicious transactions from
executing.  Our goal is to present a proof-of-concept implementation
to demonstrate the feasibility of applying the balance invariant in the
workflow of bridge transactions as a defense against malicious
transactions.  




We start by proposing a new model for the workflow of bridge
transactions.  Then, using Wormhole as an example, we show how to
modify an existing bridge implementation to prevent malicious transactions from
being processed with fewer than 40 lines of code changes.  Finally, we
evaluate the correctness and overhead of this initial implementation.

\subsection{The Report-then-Execute Model}

To guide the design of a new model, we first highlight where the
attack surfaces exist. As shown in Figure~\ref{fig:cross-chain}, a
vulnerability can exist in as early as the second step (verify deposit
amount) to as late as the next-to-last step (verify relayed
transaction). Thus, a natural place to introduce a protection
mechanism is just before executing the relayed transaction.

In our first attempt, we explored a simple mechanism that split the submission and execution of the relayed transaction into two steps. Though simple, this mechanism requires the submitter to interact with the bridge twice and increases latency. That said, this earlier iteration, which we included in the preprint version of this paper on arXiv, directly inspired Bascule~\cite{bascule}, a production project that targets a similar but distinct use case (cross-chain transfers between ETH and Bitcoin) and optimizes the process for real-world deployment.



We implement a solution similar to theirs on top of Wormhole and term this solution the \emph{report-then-execute} model. 
Figure~\ref{fig:improved-arch} depicts the entities and steps involved
in the new model, focusing on the destination chain (replacing the
right-most box of Figure~\ref{fig:cross-chain}).  The new model introduces a check step (step 12) before execution. In addition, it introduces a new role, 
\emph{Reporter}, which is responsible for asynchronously
reporting valid withdrawal transactions. Later, when the submitter submits a withdrawal transaction, the bridge checks with the reporter before execution. The reporter can be a bridge operator, or a set of trusted or trustless third-party (e.g., Bascule~\cite{bascule} uses trusted execution environments). 

In the new model, for a valid withdrawal transaction to execute, the following steps are performed: (1) the reporter asynchronously reports valid transactions that they observe (e.g., to their own contract) and announces transactions they reported through events (steps 8 and 9); (2) the submitter, upon observing that their transaction has been announced, submits the transaction; and (3) the bridge, after processing the transaction (step 11) and before executing it, consults the reporter (e.g., by calling to the reporter's contract) to verify that the transaction has been reported and the withdrawal amount satisfies the balance invariant (step 12). If the reporter confirms that the transaction is valid, the bridge executes it; otherwise, it rejects the transaction and reverts.

\subsection{Assumptions, and Trade-offs}
We assume that the reporter can independently verify the relayed transaction and the correct amount of tokens to be transferred. Additionally, we assume the bridge always calls the reporter to verify the transaction before executing it. Under this assumption, the report-then-execute model prevents any implementation bug in any existing bridge components from step 2 (e.g., the bridge fails to compute the correct deposit amount) to step 8 (e.g., the bridge fails to verify the signature of a relayed transaction). Further, if the 
reporter uses secure hardware or a decentralized third-parties
(e.g., with multisig), the model can provide additional protection against
insider threats and key theft. 

The report-then-execute model has a few advantages. Importantly, it is compatible with existing bridge designs and treats the relaying infrastructure as well as the verification code as a black box (i.e., we make no changes to these components). In addition, the submitter does not have to modify their transaction, and the withdrawal process remains one transaction. The only change is that the transaction has to be reported before submission, which can be done asynchronously and with minimal overhead in delay or latency.
As we show later below, our design integrates seamlessly into existing bridge architectures, requires minimal code changes, and only incurs a mild amount of gas overhead. 

Furthermore, our model does not introduce any new attack surfaces. Even if the reporter's key is compromised and 
reports arbitrary transactions, an attacker still has to exploit other vulnerabilities in the bridge to profit.

Third, the report-then-execute model allows any party (e.g., the bridge operator or a third-party), that has visibility into both blockchains, to determine if the
relayed transaction violates the invariant.

Lastly, our model provides an additional opportunity to reject transactions by providing a buffer between the verification and execution steps. 
In many bridge designs, if the verification step has a bug, it is not possible to stop violating transactions from being executed. In contrast, our model allows the bridge to reject transactions using the simple balance invariant, which is easy to implement and verify.

\subsection{Implementation}
\label{subsec:new-arch-impl}


Continuing our focus on Wormhole from Section~\ref{sec:live-audit}, we
modified Wormhole's withdrawal function to implement the
report-then-execute model.  We start by implementing a reporter contract that reports transactions and includes a function for checking if a transaction has been reported. We then modify Wormhole's contract. Specifically, Wormhole's current withdrawal function performs two tasks: it verifies the signatures and then executes it. We add a call to the reporter contract such that a verified transaction is checked against the reporter before execution. In total, we made fewer than 40 lines of code changes. For simplicity, we opted for a reporter with one key in our implementation. However, our implementation can be easily extended (e.g., to use multisig). 

\subsection{Evaluation}

As a final step, we performed a high-level evaluation of the
correctness and overhead (in terms of gas usage) of the
report-then-execute implementation.  We deployed our bridge
implementation on the Binance and Fantom testnets due to the
availability of free faucets for acquiring testnet tokens for them.
We then executed a series of benign and malicious transactions using
Binance as the source chain and Fantom as the destination chain.



\textbf{Correctness.}  For evaluating correctness we disabled the
security checks in the original implementation (e.g., signature
verification) to allow malicious transactions to flow through the
contract implementation (e.g., simulating key compromise).  We then
executed 100 pairs of deposit and withdrawal transactions to validate
that the implementation correctly identifies and protects itself
against malicious transactions that violate the balance invariant.

All but three of the 100 paired transactions were benign, and used
randomly generated values as transaction inputs.
The remaining three paired transactions were malicious and were
randomly placed in the transaction sequence.  The malicious
transactions represented the three kinds of bugs underlying the
large attacks in Section~\ref{sec:retro-results}: (1) a bug that
allows a user to withdraw more tokens than they deposited, (2) a bug
that allows a user to withdraw tokens without depositing any
(simulating key compromise), and (3) a bug that allows a user to
withdraw twice from the same deposit (double spending).

The bridge implementation correctly executed the benign transactions
to completion and rejected the three malicious transactions.  The
results are the same independent of where the malicious transactions
randomly appear in the sequence.

\textbf{Overhead.}  The additional steps add overhead to implementing
bridges.  For this experiment, we measured the gas usage for executing
Wormhole's original implementation (with its security checks) and the
gas usage of the report-then-execute implementation (also with the
original security checks plus our code).  On average, our implementation consumed 120\thou more gas than the
original implementation (380\thou vs.\ 270\thou gas) when
executing the benign transactions, which translates to a
roughly \$1 increase on Ethereum. 
We leave optimizations to reduce the gas overhead as future work.


\section{Related Work}


Blockchains are, by their nature, public and thus support direct
empirical analysis of past transactions.  This property has engendered a rich
literature quantifying and characterizing a range of quasi-adversarial
activities on \emph{individual} blockchains including
arbitrage~\cite{mclaughlin2023large, qin2022quantifying}, sandwich
attacks~\cite{qin2022quantifying, zhou2021high},
frontrunning~\cite{daian2020flash}, transaction
replays~\cite{qin2022quantifying}, and a range of smart contract
vulnerabilities and attacks~\cite{perez2021smart,grossman2017online,
  rodler2018sereum, zhang2020txspector, wu2021defiranger,
  ferreira2021eye}.  While our work focuses on the cross-chain
context, a number of the vulnerability classes
identified in such work are directly implicated in the attacks we analyze.




In the cross-chain context, there are several different streams of
related work.  First are the efforts to improve the security and
performance of cross-chain bridge designs --- particularly in managing
cross-chain consensus concerning state changes --- using
zero-knowledge~\cite{xie2022zkbridge} or committees of
validators~\cite{lan2021horizon,li2022polybridge}.
Some many-chain blockchain ecosystems (e.g., Avalanche) are extending
their chains with built-in bridges (across different chains). We consider our
work orthogonal to these efforts as we focus on unbalanced bridge transactions, independent of the particular
security violation that allowed such an outcome to take place.

Another line of work has reviewed real-world attacks on
cross chain
bridges~\cite{lee2023sok,zhang2023sok,notland2024sok,zhao2023comprehensive}
--- ranging from just a few events to an analysis of over 30
attacks.  Our work has directly benefited from the insight and
documentation these authors provide, but our respective focus is
distinct.  While these efforts have concentrated on identifying
the vulnerabilities and mechanisms of attack, our work is
agnostic to these details and focuses on the financial
side-effects those actions.

Yet a third research direction seeks to automatically discover new
vulnerabilities in cross-chain bridges.  Some examples of this work
use machine learning such as ChainSniper~\cite{tran2024chainsniper},
which trains models to identify vulnerable smart contracts, and Lin et
al.~\cite{lin2024detecting}, who train models to detect fake deposit
events.  Other examples use static analysis such as
XGuard~\cite{wang2024xguard} and BridgeGuard~\cite{zhou2025bridgeguard}, which statically analyzes bridge
contracts for inconsistent behaviors, and
SmartAxe~\cite{liao2024smartaxe}, which analyzes the control-flow
graph of smart contracts and identifies access control 
vulnerabilities.

The papers philosophically closest to ours are those that consider the
security properties of cross-chain bridges at a higher-level of
abstraction.  For example, Belichior et
al.~\cite{belchior2023hephaestus} build and evaluate a synthetic
bridge designed to allow high-level monitoring of bridge behavior --- including variations in financial state.  In
a more empirical context, Huang et al.~\cite{huang2024seamlessly}
characterize the transactions of the Stargate bridge, and highlight
the correlation between large or unusual trades and individual attacks. Zhang et
al.'s XScope~\cite{zhang2022xscope} and Wu et al.~\cite{wu2025safeguarding} model real-world bridge bugs using a set of
pre-defined rules and apply these to identify several attacks retrospectively. Compared to our work, the rules they consider are still considerably lower-level and more granular than our balance invariant as they are seeking to identify the cause of the detected attacks that have taken place.
Finally, perhaps the closest work to our own is the parallel work by Hu et al.~\cite{hu2024piecing}, which empirically analyzes different aspects of cross-chain bridge transactions. Our work differs in several ways. First, our work is security-oriented  and focuses exclusively on token bridges that have been attacked. In comparison, anomaly detection was only a part of Hu et al.'s objectives, while we systematically examine attack surfaces of token bridges. They identified 47 suspicious transactions in PolyNetwork and Anyswap, which is a strict subset of the 80 suspicious transactions we identified. We found more suspicious transactions because we supported a much broader set of blockchains. In addition, we flagged over a hundred more transactions in other bridges that also deserve scrutiny.
Most importantly, we propose a simple, generic defense mechanism that can be implemented in-line in existing bridge designs to prevent similar attacks in the future, which differentiates our work from any existing academic work.

Finally, a range of tools exist for 
monitoring cross-chain bridge transactions for anomalies.  These include 
Hexagate~\cite{Hexagate75:online}, Hypernative~\cite{Hypernative:online}, PeckShield~\cite{PeckShie48:online}, 
Certik~\cite{CertiKWh86:online} and ChainAegis~\cite{ChainAeg18:online} (among
others), all of which claim various levels of automated alerting for
suspicious transactions.  However, without clear information on how
these systems operate, or empirical results on their efficacy, it is hard to relate our work to theirs.

\section{Discussion and Future Work}

\label{sec:discuss}
This paper is a ``first cut'' at identifying, validating and implementing a
generic mode of theft protection into cross-chain transactions. We show that a
simple balance invariant is sufficient to detect and prevent most bridge
attacks. Our effort also highlights the need for better auditability and
auditing in the bridge ecosystem, and for new bridge designs.  

As mentioned in Section~\ref{sec:background}, in this paper we focus on
cross-chain token bridges and thus our approach does not directly apply to
bridges that have built-in exchanges and swaps.  Still, similar
invariant-checking techniques likely extend to such \emph{cross-chain exchanges}
and similar protocols (e.g., Automated Market Makers which trade between  cryptocurrencies). Such extensions will
require incorporating additional logic.

The larger opportunities opened up by this work are
more architectural.  Today, when a crypto platform declares that their
system has been audited, they typically are referring to a third-party
that inspects their smart contracts for common logic errors or
deficiencies in their process.  Any validation of overall financial
safety is inherently entwined with the totality of the implementation.
However, such code-oriented audits are inherently challenging and
ill-suited for overarching questions of financial risk.  They must
evaluate all contracts and code continuously and also ensure that no
combination of inputs or invocations might lead to a loss.  Unsurprisingly,
multiple bridges in our study had been previously audited by third parties before their attacks.

Our work suggests that by separating overall financial safety
invariants from the intricate details of every given contract or
trade, we can dramatically simplify the defender's job and their
ability to reason about risk.  Similar to so-called ``circuit
breakers'' implemented in traditional securities exchanges, the ability to script overarching financial
constraints independent of transaction details can be extremely
powerful.  For example, one might extend this work beyond a simple
cross-chain balance invariant to enforce limits on liquidity losses.
We hypothesize that by centralizing the specification and
enforcement of \emph{overall} financial constraints in one place, they
will be far easier to audit and reason about.

\newpage
\bibliographystyle{ACM-Reference-Format}
\bibliography{ref}

@article{zhang2023sok,
  title="{SoK: Security of Cross-chain Bridges: Attack Surfaces, Defenses, and Open Problems}",
  author={Zhang, Mengya and Zhang, Xiaokuan and Barbee, Josh and Zhang, Yinqian and Lin, Zhiqiang},
  journal={arXiv preprint arXiv:2312.12573},
  month = dec,
  year={2023}
}

@inproceedings{lee2023sok,
  title="{SoK: Not Quite Water Under the Bridge: Review of Cross-Chain Bridge Hacks}",
  author={Lee, Sung-Shine and Murashkin, Alexandr and Derka, Martin and Gorzny, Jan},
  booktitle={Proceedings of the 2023 IEEE International Conference on Blockchain and Cryptocurrency (ICBC)},
  pages={1--14},
  year={2023},
  month = may,
  organization={IEEE}
}

@article{zhao2023comprehensive,
  title="{A Comprehensive Overview of Security Vulnerability Penetration Methods in Blockchain Cross-Chain Bridges}",
  author={Zhao, Qianrui and Wang, Yinan and Yang, Bo and Shang, Ke and Sun, Maozeng and Wang, Haijun and Yang, Zijiang and Xin, Haojie},
  journal={Authorea Preprints},
  month = oct,
  year={2023},
  publisher={Authorea}
}

@inproceedings{zhang2022xscope,
  title="{Xscope: Hunting for Cross-Chain Bridge Attacks}",
  author={Zhang, Jiashuo and Gao, Jianbo and Li, Yue and Chen, Ziming and Guan, Zhi and Chen, Zhong},
  booktitle={Proceedings of the 37th IEEE/ACM International Conference on Automated Software Engineering (ASE)},
  pages={1--4},
  year={2022},
  month = oct,
}

@misc{wormhole,
author = {},
title = {Wormhole Bridge},
howpublished = {\url{https://wormhole.com}},
month = {03},
year = {2024},
note = {Accessed 2024-08-28}
}

@misc{wormholeattack,
author = {CertiK},
title = "{Wormhole Bridge Exploit Incident Analysis}",
howpublished = {\url{https://www.certik.com/resources/blog/1kDYgyBcisoD2EqiBpHE5l-wormhole-bridge-} \url{exploit-incident-analysis}},
month = jul,
year = {2022},
}

@online{nomad,
author = {},
title = {Nomad | Bridge},
url = {https://app.nomad.xyz/},
month = {03},
year = {2024},
urldate = {03/23/2024}
}

@online{ronin,
author = {},
title = {Bridge | Ronin App},
url = {https://app.roninchain.com/},
month = {03},
year = {2024},
urldate = {03/23/2024}
}

@misc{roninattack,
author = {Andrew Thurman},
title = "{Axie Infinity's Ronin Network Suffers \$625M Exploit}",
howpublished = {\url{https://www.coindesk.com/tech/2022/03/29/axie-infinitys-ronin-} \url{network-suffers-625m-exploit/}},
month = mar,
year = {2022},
}

@misc{GithubBridgeBugTracker,
author = {0xDatapunk},
title = {GitHub -- 0xDatapunk/Bridge-Bug-Tracker},
howpublished = {\url{https://github.com/0xDatapunk/Bridge-Bug-} \url{Tracker}},
month = {},
year = {},
note = {Accessed 2024-08-28}
}

@misc{SlowMistHackedBridges:online,
key = {SlowMist},
title = "{SlowMist Hacked -- SlowMist Zone}",
howpublished = {\url{https://hacked.slowmist.io/?c=Bridge}},
month = {},
year = {},
note = {Accessed 2024-08-28}
}

@misc{REKTDB:online,
key = {defi},
title = "{De.Fi REKT Database}",
howpublished = {\url{https://de.fi/rekt-} \url{database}},
month = {},
year = {},
note = {Accessed 2024-08-28}
}

@misc{Web3Great:online,
author = {Molly White},
title = "{Web3 is Going Just Great}",
howpublished = {\url{https://www.web3isgoinggreat.com/charts/top}},
month = {},
year = {},
note = {Accessed 2024-08-28}
}

@misc{GithubBridgeHacks2:online,
author = {Chris Whinfrey},
title = "{Bridge Hack List -- GitHub}",
howpublished = {\url{https://gist.github.com/cwhinfrey/9fd1bbc31bbcff08fca242b90c7f875d}},
month = {},
year = {},
note = {Accessed 2024-07-02}
}

@misc{Finality:online,
author = {Georgios Konstantopoulos and Vitalik Buterin},
title = "{Ethereum Reorgs After The Merge}",
howpublished = {\url{https://www.paradigm.xyz/2021/07/ethereum-reorgs-after-} \url{the-merge}},
month = jul,
year = 2021,
}

@misc{defillama-volume,
author = {DefiLlama},
title = "{Bridge Volume}",
howpublished = {\url{https://defillama.com/bridges}},
note = {Accessed 2024-08-28},
}

@inproceedings{mclaughlin2023large,
  title="{A Large Scale Study of the Ethereum Arbitrage Ecosystem}",
  author={McLaughlin, Robert and Kruegel, Christopher and Vigna, Giovanni},
  booktitle={Proceedings of the 32nd USENIX Security Symposium (USENIX Security)},
  pages={3295--3312},
  year={2023},
  month = aug,
  organization = {USENIX},
}

@inproceedings{qin2022quantifying,
  title="{Quantifying Blockchain Extractable Value: How dark is the forest?}",
  author={Qin, Kaihua and Zhou, Liyi and Gervais, Arthur},
  booktitle={Proceedings of the 2022 IEEE Symposium on Security and Privacy (S\&P)},
  pages={198--214},
  year={2022},
  month = may,
  organization={IEEE}
}

@inproceedings{zhou2021high,
  title="{High-Frequency Trading on Decentralized On-Chain Exchanges}",
  author={Zhou, Liyi and Qin, Kaihua and Torres, Christof Ferreira and Le, Duc V and Gervais, Arthur},
  booktitle={Proceedings of the 2021 IEEE Symposium on Security and Privacy (S\&P)},
  pages={428--445},
  year={2021},
  month = may,
  organization={IEEE}
}

@inproceedings{zhang2020txspector,
  title="{TxSpector: Uncovering Attacks in Ethereum from Transactions}",
  author={Zhang, Mengya and Zhang, Xiaokuan and Zhang, Yinqian and Lin, Zhiqiang},
  booktitle={Proceedings of the 29th USENIX Security Symposium (USENIX Security)},
  pages={2775--2792},
  year={2020},
  month = aug,
}

@article{wu2021defiranger,
  title="{DeFiRanger: Detecting DeFi Price Manipulation Attacks}",
  author={Wu, Siwei and Yu, Zhou and Wang, Dabao and Zhou, Yajin and Wu, Lei and Wang, Haoyu and Yuan, Xingliang},
  journal={IEEE Transactions on Dependable and Secure Computing},
  year={2024},
  month = "July/August",
  volume = 21,
  number = 4,
  pages = "4147-4161",
}

@inproceedings{ferreira2021eye,
  title="{The Eye of Horus: Spotting and Analyzing Attacks on Ethereum Smart Contracts}",
  author={Ferreira Torres, Christof and Iannillo, Antonio Ken and Gervais, Arthur and State, Radu},
  booktitle={Proceedings of the 25th International Conference on Financial Cryptography and Data Security (FC)},
  pages={33--52},
  year={2021},
  month = mar,
  organization={Springer}
}

@inproceedings{daian2020flash,
  title="{Flash boys 2.0: Frontrunning in Decentralized Exchanges, Miner Extractable Value, and Consensus Instability}",
  author={Daian, Philip and Goldfeder, Steven and Kell, Tyler and Li, Yunqi and Zhao, Xueyuan and Bentov, Iddo and Breidenbach, Lorenz and Juels, Ari},
  booktitle={Proceedings of the 2020 IEEE Symposium on Security and Privacy (S\&P)},
  pages={910--927},
  year={2020},
  month = may,
  organization={IEEE}
}

@article{grossman2017online,
  title="{Online Detection of Effectively Callback Free Objects with Applications to Smart Contracts}",
  author={Grossman, Shelly and Abraham, Ittai and Golan-Gueta, Guy and Michalevsky, Yan and Rinetzky, Noam and Sagiv, Mooly and Zohar, Yoni},
  journal={Proceedings of the ACM on Programming Languages},
  volume={2},
  number={POPL},
  pages={1--28},
  year={2018},
  publisher={ACM}
}

@inproceedings{perez2021smart,
  title="{Smart Contract Vulnerabilities: Vulnerable Does Not Imply Exploited}",
  author={Perez, Daniel and Livshits, Benjamin},
  booktitle={Proceedings of the 30th USENIX Security Symposium (USENIX Security)},
  pages={1325--1341},
  year={2021},
  month = aug,
}

@inproceedings{xie2022zkbridge,
  title="{zkBridge: Trustless Cross-chain Bridges Made Practical}",
  author={Xie, Tiancheng and Zhang, Jiaheng and Cheng, Zerui and Zhang, Fan and Zhang, Yupeng and Jia, Yongzheng and Boneh, Dan and Song, Dawn},
  booktitle={Proceedings of the 2022 ACM SIGSAC Conference on Computer and Communications Security (CCS)},
  pages={3003--3017},
  year={2022},
  month = nov,
}

@article{lan2021horizon,
  title="{Horizon: A Gas-Efficient, Trustless Bridge for Cross-Chain Transactions}",
  author={Lan, Rongjian and Upadhyaya, Ganesha and Tse, Stephen and Zamani, Mahdi},
  journal={arXiv preprint arXiv:2101.06000},
  year={2021},
  month = jan,
}

@inproceedings{li2022polybridge,
  title="{POLYBRIDGE: A Crosschain Bridge for Heterogeneous Blockchains}",
  author={Li, Yue and Liu, Han and Tan, Yuan},
  booktitle={2022 IEEE International Conference on Blockchain and Cryptocurrency (ICBC)},
  pages={1--2},
  year={2022},
  month = may,
  organization={IEEE}
}

@article{rodler2018sereum,
  title="{Sereum: Protecting Existing Smart Contracts Against Re-Entrancy Attacks}",
  author={Rodler, Michael and Li, Wenting and Karame, Ghassan O and Davi, Lucas},
  journal={arXiv preprint arXiv:1812.05934},
  year={2018},
  month = dec,
}

@inproceedings{huang2024seamlessly,
  title="{Seamlessly Transferring Assets through Layer-0 Bridges: An Empirical Analysis of Stargate Bridge's Architecture and Dynamics}",
  author={Huang, Chuanshan and Yan, Tao and Tessone, Claudio J},
  booktitle={Proceedings of the 2024 Web Conference},
  pages={1776--1784},
  year={2024},
  month = may,
}

@article{belchior2023hephaestus,
  title="{Hephaestus: Modeling, Analysis, and Performance Evaluation of Cross-Chain Transactions}",
  author={Belchior, Rafael and Somogyvari, Peter and Pfannschmidt, Jonas and Vasconcelos, Andr{\'e} and Correia, Miguel},
  journal={IEEE Transactions on Reliability},
  month = jun,
  volume = {73},
  number = {2},
  year={2023},
  publisher={IEEE},
  pages = {1132-1146},
}

@article{notland2024sok,
  title="{SoK: Cross-Chain Bridging Architectural Design Flaws and Mitigations}",
  author={Notland, Jakob Svennevik and Li, Jinguye and Nowostawski, Mariusz and Haro, Peter Halland},
  journal={arXiv preprint arXiv:2403.00405},
  year={2024},
  month = mar,
}

@inproceedings{tran2024chainsniper,
  title="{ChainSniper: A Machine Learning Approach for Auditing Cross-Chain Smart Contracts}",
  author={Tran, Tuan-Dung and Vo, Kiet Anh and Phan, Duy The and Tan, Cam Nguyen and Pham, Van-Hau},
  booktitle={Proceedings of the 9th International Conference on Intelligent Information Technology (ICIIT)},
  pages={223--230},
  year={2024},
  month = feb,
}

@inproceedings{lin2024detecting,
  title="{Detecting Fake Deposit Attacks on Cross-chain Bridges from a Network Perspective}",
  author={Lin, Kaixin and Lin, Dan and Zheng, Ziye and Tan, Yixiang and Wu, Jiajing},
  booktitle={Proceedings of the 2024 IEEE International Symposium on Circuits and Systems (ISCAS)},
  pages={1--5},
  year={2024},
  month = may,
  organization={IEEE}
}

@inproceedings{wang2024xguard,
  title="{XGuard: Detecting Inconsistency Behaviors of Crosschain Bridges}",
  author={Wang, Ke and Li, Yue and Wang, Che and Gao, Jianbo and Guan, Zhi and Chen, Zhong},
  booktitle={Proceedings of the 32nd ACM International Conference on the Foundations of Software Engineering (FSE)},
  pages={612--616},
  year={2024},
  month = jul,
}

@article{liao2024smartaxe,
  title="{SmartAxe: Detecting Cross-Chain Vulnerabilities in Bridge Smart Contracts via Fine-Grained Static Analysis}",
  author={Liao, Zeqin and Nan, Yuhong and Liang, Henglong and Hao, Sicheng and Zhai, Juan and Wu, Jiajing and Zheng, Zibin},
  journal={Proceedings of the ACM on Software Engineering},
  volume={1},
  number={FSE},
  pages={249--270},
  year={2024},
  month = jul,
  publisher={ACM}
}

@misc{nomad-groundtruth-github:online,
author = {Nomad},
title = "{Nomad Hack Data -- Github}",
howpublished = {\url{https://github.com/nomad-xyz/hack-data}},
note = {Accessed 2024-09-04}
}

@misc{meter-groundtruth-tencent:online,
author = {PANews},
title = "{Another cross-chain bridge project was attacked, and Meter.io lost \$4.2 million}",
howpublished = {\url{https://new-qq-com.translate.goog/rain/a/20220206A05XNI00?_x_tr_sl=zh-CN&_x_tr_tl=en&_x_tr_hl=en&_x_tr_pto=sc}},
note = {Accessed 2024-09-04}
}

@misc{chainswap-groundtruth:online,
author = {Liquid},
title = "{ChainSwap -- GitHub}",
howpublished = {\url{https://github.com/liqtags/crypto-rekts/blob/main/rekts/} \url{ChainSwap-2.md}},
note = {Accessed 2024-09-04}
}

@misc{qubit:rekt,
author = {rekt},
title = "{Qubit Finance -- REKT}",
howpublished = {\url{https://rekt.news/qubit-rekt/}},
month = {},
year = {},
note = {Accessed 2024-09-04}
}

@misc{anyswapattack,
  title = {Multichain Hack Worsens as Loss of Funds Reaches {\textdollar 3M}: Report},
  author = {{CoinDesk}},
  year = {2022},
  howpublished = {\url{www.coindesk.com/business/2022/01/20/multichain-hack-worsens-as-loss-of-funds-reaches-3m-report/}},
  note = {Accessed: 2024-09-04}
}

@online{binanceproof:online,
author = {emiliano},
title = {emiliano on X: they used a proof from one of the very first cross chain tx},
url = {https://x.com/emilianobonassi/status/1578742891538567169},
month = {09},
year = {2024},
urldate = {09/04/2024}
}

@online{pNetworkhack:online,
author = {pNetwork Team},
title = {pNetwork Post Mortem: pBTC-on-BSC Exploit},
howpublished = {\url{https://medium.com/pnetwork/pnetwork-post-mortem-pbtc-on-bsc-exploit-170890c58d5f}},
month = {11},
year = {2024},
urldate = {11/13/2024}
}

@online{Thorhack1:online,
author = {THORChain},
title = {Post-mortem: ETH Router Exploits 1 \& 2, and premature Return To Trading Incident},
howpublished = {\url{https://medium.com/thorchain/post-mortem-eth-router-exploits-1-2-and-premature-return-to-trading-incident-2908928c5fb}},
month = {11},
year = {2024},
urldate = {11/13/2024}
}

@online{Thorhack2:online,
author = {THORChain},
title = {THORChain Incident 07.15},
howpublished = {\url{https://thearchitect.notion.site/THORChain-Incident-07-15-7d205f91924e44a5b6499b6df5f6c210}},
month = {11},
year = {2024},
urldate = {11/13/2024}
}

@online{evoDefihack:online,
author = {Tom Carreras and Mike Dalton},
title = {EVODeFi Bridge May Be Missing \$66M in Funds},
howpublished = {\url{https://cryptobriefing.com/oasis-evodefi-bridge-may-be-missing-66m-in-funds/}},
month = {11},
year = {2024},
urldate = {11/13/2024}
}

@online{Multichainhack:online,
author = {Multichain},
title = {Multichain (Previously Anyswap) on X},
howpublished = {\url{https://x.com/multichainorg/status/1677096839731097600}},
month = {11},
year = {2024},
urldate = {11/13/2024}
}

@online{Hexagate75:online,
author = {Hexagate},
title = {Hexagate},
howpublished = {\url{https://www.hexagate.com/real-time-prevention}},
month = {11},
year = {2024},
urldate = {11/13/2024}
}

@online{Hypernative:online,
author = {Hypernative},
title = {Hypernative: Web3 Security for Chains},
howpublished = {\url{https://www.hypernative.io/solutions/chains}},
month = {11},
year = {2024},
urldate = {11/13/2024}
}

@online{CertiKWh86:online,
author = {CertiK},
title = {CertiK},
howpublished = {\url{https://www.certik.com/resources/blog/what-is-on-chain-monitoring}},
month = {11},
year = {2024},
urldate = {11/13/2024}
}

@online{PeckShie48:online,
author = {PeckShield},
title = {PeckShield Alert},
howpublished = {\url{https://alert.peckshield.com/}},
month = {11},
year = {2024},
urldate = {11/13/2024}
}

@online{ChainAeg18:online,
author = {ChainAegis},
title = {ChainAegis - The World's Leading AI-driven risk detection and alerting platform},
howpublished = {\url{https://www.chainaegis.com/}},
month = {11},
year = {2024},
urldate = {11/13/2024}
}

@online{bascule,
author = {Cubist},
title = {Cubist on X --- Bascule Drawbridge},
howpublished = {\url{https://x.com/cubistdev/status/1849093545636036832}},
month = {11},
year = {2024},
urldate = {11/14/2024}
}

@article{hu2024piecing,
  title={Piecing Together the Jigsaw Puzzle of Transactions on Heterogeneous Blockchain Networks},
  author={Hu, Xiaohui and Feng, Hang and Xia, Pengcheng and Tyson, Gareth and Wu, Lei and Zhou, Yajin and Wang, Haoyu},
  journal={Proceedings of the ACM on Measurement and Analysis of Computing Systems},
  volume={8},
  number={3},
  pages={1--27},
  year={2024},
  publisher={ACM New York, NY, USA}
}

@inproceedings{wu2025safeguarding,
  title={Safeguarding blockchain ecosystem: Understanding and detecting attack transactions on cross-chain bridges},
  author={Wu, Jiajing and Lin, Kaixin and Lin, Dan and Zhang, Bozhao and Wu, Zhiying and Su, Jianzhong},
  booktitle={Proceedings of the ACM on Web Conference 2025},
  pages={4902--4912},
  year={2025}
}

@article{zhou2025bridgeguard,
  title={BridgeGuard: Checking External Interaction Vulnerabilities in Cross-Chain Bridge Router Contracts Based on Symbolic Dataflow Analysis},
  author={Zhou, Zequan and Luo, Xiling and Ji, Xiaohai and Mao, Jian and He, Ting and Wang, Junjun and Wu, Qianhong},
  journal={IEEE Transactions on Dependable and Secure Computing},
  year={2025},
  publisher={IEEE}
}
\clearpage
\appendix
\section{Ethics}
\label{sec:ethics}

We believe our work, which deals with public data, no identified
individuals and a simple means for identifying attacks on crypto token
transfer bridges (and potentially preventing such attacks in the
future), has very low ethical risk and significant upside. Moreover, we have attempted to disclose significant suspicious bridge
transactions to appropriate bridge operators (yet with limited
success as most of the bridges have ceased operations). Lastly, we will make our code and data available upon publication.

\section{Attack Descriptions and Additional Details}
\label{sec:attack-summaries}

In this section we provide a descriptive account of each of the
attacks listed in Table~\ref{table:bridges-and-txns-flagged} in
Section~\ref{sec:retro-results}, giving additional background on the
bridge, how the attack transpired, and more details on our results
using \offlinetool.

\subsection{Ronin Bridge}
\textbf{Background.} Ronin bridge operates between Ethereum and the Ronin chain. It was hacked in March 2022. The attacker compromised the bridge's private keys, allowing them to mint arbitrary amounts of assets. The attacker carried out two transactions, minting around \$624 million USD worth of assets on Ethereum.

\pgraph{Results.} \offlinetool~analyzed over 3 million bridge transactions and alerted on both attack transactions.

\subsection{Poly Network Bridge (2021)}
\textbf{Background.} PolyNetwork is a cross-chain bridge that supports asset transfers between multiple blockchains (e.g., BSC, ETH, and Polygon). It was hacked on August 10th, 2021. The attacker exploited a bug in the bridge's verification code, allowing them to insert their own keys and verify any malicious payload. Overall, the attacker stole around \$600 million USD worth of assets on BSC, ETH, and Polygon.

\pgraph{Results.} In total, \offlinetool~analyzed over 292\thou bridge transactions between ETH, BSC, Polygon and Poly Network's liquidity pool. \offlinetool~alerted on all 18 bridge transactions in the attack. In addition, \offlinetool~also flagged one withdrawal that seemingly was relayed to the wrong destination chain.

\subsection{Binance Token Hub}
\textbf{Background.} Binance Token Hub
facilitates asset transfers between Binance Beacon Chain and Binance Smart Chain. It was hacked on
October 7th, 2022. The attacker exploited a bug in Binance Bridge's verification code, allowing them to verify carefully crafted malicious payloads. The attacker carried out the attack in two transactions, each minting 1,000,000 BNB on Binance Smart Chain (approximately \$580 million USD in total).

\pgraph{Results.} \offlinetool~analyzed over 2M bridge transactions between 
Binance Beacon Chain and Binance Smart Chain. 
\offlinetool~only alerted on the two attack transactions, citing the discrepancy between the amount given out and the amount received by the bridge. We end by noting that the partner deposit transactions returned by the Binance Token Hub's API for two attack transactions are different than the ones suggested in some blogposts~\cite{binanceproof:online}. However, the attack transactions will be flagged regardless of which partner deposit transactions is used.

\subsection{Wormhole}
\textbf{Background.} 
Wormhole is a general-purpose cross-chain bridge that 
currently supports around 35 blockchains. It was hacked on February 2nd, 2022. Specifically, the attacker exploited a bug in Wormhole's smart contract on Solana that allowed them to verify arbitrary unauthorized payload. The attacker executed one transaction and minted 120,000 wETH (around \$350 million) on Solana.

\pgraph{Results.} 
In total, \offlinetool~analyzed over 642k transactions across ten blockchains. It alerted on three transactions, one of which was the attack transaction.

\subsection{Nomad Bridge}
\textbf{Background.} Nomad bridge support asset transfers across six blockchains. It was hacked on August 1st, 2022. The attacker exploited a bug in the bridge's verification code, allowing them to verified any payload that had not been verified before. Shortly after the first a few attack transactions, a group of copycats joined the crusade of draining the bridge. In total, the reported loss was around \$190 million.

\pgraph{Results.} \offlinetool~analyzed over 37k transactions, and alerted on 962 transactions that transferred assets to 561 unique addresses. We found one dataset that reported 561 addresses, which matched exactly with the addresses we identified. We also found one Github repository that mentioned identifying 960 transactions~\cite{nomad-groundtruth-github:online}. As mentioned earlier, we identified two more transactions. Upon manually inspection, we confirm that the two transactions were indeed malicious.

\subsection{Harmony Bridge}
\textbf{Background.} Harmony bridge operates between ETH, BSC and Harmony. It was hacked on June 24, 2022. The attacker compromised two of the signing keys of the bridge, allowing them to mint arbitrary amounts of assets. In total, the attacker minted around \$100 million worth of assets on BSC and ETH in 15 transactions.

\pgraph{Results.} \offlinetool~analyzed over 336k transactions and alerted on 58 transactions, including all 15 attack transactions.

\subsection{HECO Bridge}
\textbf{Background.} HECO bridge allows users to transfer assets between Huobi ECO Chain (HECO) and Ethereum. It was hacked on November 11th, 2023. The attacker compromised the bridge's private keys, allowing them to sign arbitrary transactions. The attacker carried out eight transactions, minting around \$86 million worth of assets on Ethereum.

\pgraph{Results.} \offlinetool~analyzed over 23k transactions. All eight attack transactions were flagged by \offlinetool.

\subsection{Qubit Bridge}
\textbf{Background.} Qubit bridge allows users to transfer assets between ETH and BSC. It was hacked on January 27th, 2022. The attacker exploited a bug in the deposit function, which allowed them to trick the bridge into believing that a deposit had been made when it had not. The attacker carried out 16 transactions, stealing around \$80 million worth of assets.

\pgraph{Results.} \offlinetool~analyzed over 260 transactions and alerted on all 16 transactions. 


\subsection{Anyswap Bridge}
\textbf{Background.} Anyswap bridge supported moving assets across many blockchains at the time of the attack. It was hacked on July 10, 2021. The attacker exploited a bug in the bridge's verification code, allowing them to verify any malicious payload. The attacker carried out three transactions, minting around \$7.9\mil worth of assets on Ethereum.


\pgraph{Results.}
In addition to the previously reported transactions from the July 10, 2021,
attack, Anyswap had more than 800 additional transactions that violated the
balance invariant.

\subsection{Poly Network Bridge (2023)}
\textbf{Background.} Shortly after the hack in 2021,
PolyNetwork switched to a new set of smart contracts.  It was,
however, hacked again in August 10th, 2021. The attacker exploited a
bug in the bridge's verification code, allowing them to verify
arbitrary payload. Overall, there were 136 reported transactions.

\pgraph{Results.} \offlinetool~analyzed over 290k transactions and flagged all 136 transactions. \offlinetool~also flagged 27 transactions that pointed to the same non-existent deposit transaction hash (\hash{0x0101...} with 01 repeated 32 times).
All of these transactions occurred on Aug 22, 2023 and withdrew funds from the bridges without backing deposits. In total, over \$20 million worth of assets were withdrawn in these transactions.

\subsection{Chainswap Bridge}
\textbf{Background.} Chainswap bridge support token transfers between five bridges and 
was hacked on July 10, 2021. The attacker exploited a bug in the bridge's verification code, allowing them to verify any malicious payload. The attacker stole \$4.4 million worth of assets on Ethereum and BSC using one address. Of particular note, unlike other bridges, Chainswap did not publicly disclose the list of bridge contract addresses. 
As such, we programmatically identified all potential bridge contract addresses by searching for the specific event signatures that known Chainswap bridge contracts emit.

\pgraph{Results.} \offlinetool~analyzed over 53k transactions. \offlinetool~alerted on 1136 transactions, all of which were executed by the single malicious address. We note that we are unable to find any public information on the list of transactions that were part of the attack, preventing us from verifying individual transactions.

\subsection{Meter Bridge}
\textbf{Background.} Meter bridge allows users to transfer between Meter's own chain and a few EVM-based chains. It was hacked on February 5, 2022. The attacker exploited a bug in the bridge's deposit function, where the attacker tricked the bridge into believing that a deposit had been made. In total, the attacker carried out 5 transactions, stealing around \$4.3 million worth of assets.

\pgraph{Results.} \offlinetool~analyzed over 14k transactions, alerting on all five attack transactions.


\section{Additional Transactions Identified}

Table~\ref{tab:xaction-hashes} provides additional details on specific
example bridge transactions that violate the balance invariant and
have otherwise not been previously reported.  For each example, the
table lists the hashes of the paired deposit and withdrawal
transactions, the blockchain, and the number of tokens transferred. If
a deposit transaction does not exist or is otherwise invalid, we mark
its chain and token as N/A.
\clearpage

\begin{table*}[t!]
\centering
\scriptsize
\begin{tabular}{p{1.2cm}p{4.3cm}cp{4.3cm}p{1.3cm}}
\toprule
\textbf{Bridge} & \textbf{\makecell{(Claimed) Deposit \\Transaction Hash}} & \textbf{\makecell{Chain \&\\Token}} & \textbf{Withdraw Transaction Hash} & \textbf{\makecell{Chain \\\& Token}} \\
\midrule
Wormhole \#1 & \hash{0x8bbb7befd198a5e90297f451fc43a9e90de083289a041c8af94116c785cf496d} &\makecell{Polygon\\0.5\\WSOL}  & \hash{5AiesW9pKrZvCCJM8QPWmqx...dkadV1waPVLWfsnCMVQmyYaciLxEo} &\makecell{SOL\\650\\USDC}\\[0.2in]
Wormhole \#2 & \hash{0x55d1e486a8e2102e07fd6270a03f05bbee7b43bf27ebac97b95b98e068f6740e} &\makecell{Polygon\\4.3k\\MATIC}  & \hash{0xbe81895b1c3172fd69b8d4d9bf726edfdc17083876c440f9414ff316999237d7} &\makecell{AVAX\\2.3k\\WMATIC}\\

\hline

Anyswap \#1 & \hash{0x01ba4719c80b6fe911b091a7c05124b64eeece964e09c058ef8f9805daca546b} & N/A &  \hash{0xf015a6b06a13a08d3499ece17504d14a95d6af3e04ae11f291dca22dbbf6c991}  & \makecell{BSC\\$5*10^{-8}$\\USDC}\\[0.2in]
Anyswap \#2 & \hash{0x01ba4719c80b6fe911b091a7c05124b64eeece964e09c058ef8f9805daca546b} & N/A &  \hash{0x9e55b7295880dce76aa8af0f3e3f9e36499ae0bdb28088a5924daf29c6132ceb}  & \makecell{BSC\\55k\\USDC} \\[0.2in]

Anyswap \#3 & \hash{0xe3b0c44298fc1c149afbf4c8996fb92427ae41e4649b934ca495991b7852b855} & N/A &  \hash{0x4f038804d0622d2eab15d21d902a3fdd3bdfb5427bb5fd65b9eb0a41169534be}  & \makecell{Polygon\\100k\\USDC}\\[0.2in]

Anyswap \#4 & \hash{0x0x0000000000000000000000000000000000000000000000000000000000000000} & N/A &  \hash{0x98aa9e94d4fd0a05c27eb13ac2e699e4426c8dd9d57d04c0fa09cf4951eb2f94}  & \makecell{BSC\\650k\\USDC}\\[0.2in]

Anyswap \#5 & \hash{0x0x0000000000000000000000000000000000000000000000000000000000000000} & N/A &  \hash{0xa67ac5dc308142f89409df89dc85e8fab88c575b3adef77fbc8f51858b7bf7cb}  & \makecell{Polygon\\50k\\USDC}\\[0.2in]

Anyswap \#6 & \hash{0x28b233a4dbda8b4dfae7245b8fff434de95f6dbd101e1a9cb22a95ded1315a16} & \makecell{Fantom\\6.4k\\POPS} & \hash{0x76bdcfd5ddfa358bf4181556e3b4f1fdd2d648a246bfab91386bdfbd7b76d01f} & \makecell{Avalanche\\6.4k\\POPS}\\[0.2in]

& \hash{0xf0b5568dfd8a4559d30adc9dfc881875210a3b9dfa680d392b33eb1d2cc86cfa} & \makecell{Fantom\\6.4k\\POPS} & \hash{0xc86297f14f32a33232149025d4e8f8e50985d76ac1b7ccaf181501820c0b1cf7} & \makecell{Avalanche\\6.4k\\(any)POPS}\\[0.2in]

Anyswap \#7 & \hash{0x0x0000000000000000000000000000000000000000000000000000000000000000} & N/A &  \hash{0xde790e8dc59d8bae7ebdf89c4b75267a6e0783219b32aebe83e112aac6c299f5}  & \makecell{Avalanche\\54k\\USDC}\\

\hline
HECO \#1 & \hash{0x6f9d2e82aef87fc649198976974c05d4c540dacca5043ffee619cc33f3ba4cf5} & \makecell{ETH\\5m\\USDT} & \hash{0x628e878fb723cf0dd838eb956ce78d23b45b130876a625fd4d283e62ac2289f0}  & \makecell{HECO\\5m\\USDT}\\[0.2in]

& \hash{0x6f9d2e82aef87fc649198976974c05d4c540dacca5043ffee619cc33f3ba4cf5} & \makecell{ETH\\5m\\USDT} & \hash{0x27a1e6a66b6e0fc5fa805f7400dd07397bb92226926868a82afb44154a32128b} & \makecell{HECO\\5m\\USDT}\\

\hline
Harmony \#1 & \hash{0x559bc92656a6956a5ffe9eea6f14a5d5993520e31a1a08551d5171ad8f658886} & \makecell{BSC\\5.3k\\BUSD} & \hash{0xdf3bf1a8227ede87d7905c026c3b6a3504cc81399ebd08e1273e1a9dd2c748a9}  & \makecell{Harmony\\5.3k\\BUSD}\\[0.2in]

& \hash{0x559bc92656a6956a5ffe9eea6f14a5d5993520e31a1a08551d5171ad8f658886} & \makecell{BSC\\5.3k\\BUSD} & \hash{0x304801a2b33585e6867de0c403535588979ce4d2cf41c6922223d3203589c39d} & \makecell{ETH\\$5*10^{-18}$\\BUSD}\\ 

\hline
PolyNet. \#1 & \hash{0x0101010101010101010101010101010101010101010101010101010101010101} & N/A &  \hash{0xd6b7f50e974311082eb4b413219f7198cbf897af4e0f2e9202b10c6afe8fa0a2}  & \makecell{ETH\\491\mil\\PLT}\\ 

\bottomrule
\end{tabular}
\caption{Other flagged transactions. PolyNet stands for Poly Network 2023. Full transaction hash for Solana: \protect\hash{5AiesW9pKrZvCCJM8QPWmqxsnRoQwHaQmX8NR9a8BFz3pmt2ypW67zgqeRWdkadV1waPVLWfsnCMVQmyYaciLxEo}.}
\label{tab:xaction-hashes}
\end{table*}

\end{document}